\documentclass[aps,prl,twocolumn,amsmath,amssymb,superscriptaddress,floatfix]{revtex4-2}

\usepackage{mymacros}

\begin{document}
\title{Large-momentum-transfer atom interferometers with \texorpdfstring{$\mu$}{u}rad-accuracy using Bragg diffraction}

\def\affitp  {\affiliation{Leibniz Universit\"at Hannover, Institut f\"ur Theoretische Physik, Appelstr. 2, D-30167 Hannover, Germany }}
\def\affiqo  {\affiliation{Leibniz Universit\"at Hannover, Institut f\"ur Quantenoptik, Welfengarten 1, D-30167 Hannover, Germany }}
\def\affdlr  {\affiliation{Deutsches Zentrum f\"ur Luft- und Raumfahrt e. V. (DLR),\\
Institut f\"ur Satellitengeod\"asie und Inertialsensorik, Callinstraße 30b, D-30167 Hannover, Germany}}
\author{J.-N.~Siem\ss}
\email[email (he/him/his): ]{jan-niclas.siemss@itp.uni-hannover.de}
\affitp
\affiqo
\author{F.~Fitzek}
\affitp
\affiqo
\author{C.~Schubert}
\affiqo
\affdlr 
\author{E.~M.~Rasel}
\author{N.~Gaaloul}
\affiqo
\author{K.~Hammerer}
\email[email: ]{klemens.hammerer@itp.uni-hannover.de}
 \affitp
\date{\today}

\begin{abstract}
Large-momentum-transfer~(LMT) atom interferometers using elastic Bragg scattering on light waves are among the most precise quantum sensors to date. To advance their accuracy from the mrad to the $\mu$rad regime, it is necessary to understand the rich phenomenology of the Bragg interferometer, which differs significantly from that of a standard two-mode interferometer. We develop an analytic model for the interferometer signal and demonstrate its accuracy using comprehensive numerical simulations. Our analytic treatment allows the determination of the atomic projection noise limit of an LMT Bragg interferometer, and provides the means to saturate this limit. It affords accurate knowledge of the systematic phase errors as well as their suppression by two orders of magnitude down to a few $\mu\mathrm{rad}$ using appropriate light pulse parameters.
\end{abstract}
\maketitle

%%%%%%%%%%%%%%%%%%%%%%%%%%%%%%%%%%%%%%%%%%%%%%%%%%%%%%%%%%%%%%%%
%%%     Section: Introduction
%%%%%%%%%%%%%%%%%%%%%%%%%%%%%%%%%%%%%%%%%%%%%%%%%%%%%%%%%%%%%%%%
Atom interferometery enables the most precise determination of the fine-structure constant~\cite{Parker2018,Morel2020} as well as the most accurate quantum test of the universality of free fall~\cite{Asenbaum2020}. The unique ability to perform absolute measurements of inertial forces~\cite{Geiger2020} with high accuracy and precision makes atom interferometers prime candidates for real-world applications~\cite{Bongs2019} like gravimetry~\cite{Menoret2018,Wu2019}, gravity cartography~\cite{Stray2022}, and inertial navigation~\cite{Geiger2011,Cheiney2018}. It has also recently led to the first measurement of the gravitational Aharonov-Bohm effect~\cite{Overstreet2022}.
Large-momentum-transfer~(LMT) beam splitting exploits the enhanced scaling of the interferometer sensitivity with the spatial separation of the coherent superposition of matter waves. Several proof-of-principle experiments have demonstrated record-breaking separations~\cite{Chiow2011,Plotkin-Swing2018,Gebbe2021,Wilkason2022} using various beam splitting techniques. LMT holds the potential for unprecedented precision of quantum sensors and will greatly facilitate our understanding of fundamental physics, e.g., through the detection of gravitational waves as well as the search for ultra-light dark matter~\cite{Graham2013,Canuel2018,Canuel2020,Zhan2020,Badurina2020,Abe2021}.

\begin{figure}[t]
        \includegraphics[width=\columnwidth]{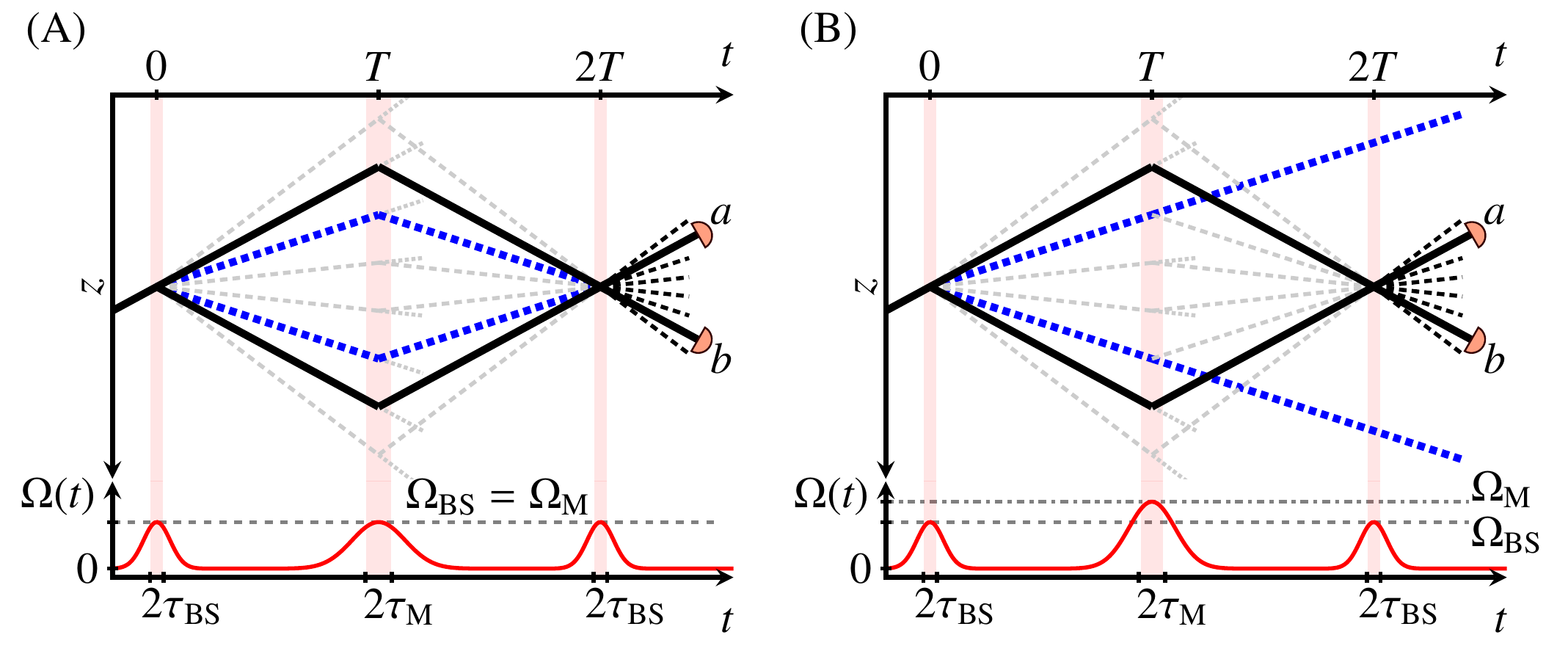}
        \caption{Mach-Zehnder~(MZ) interferometer (solid trajectories) realized by $\BraggOrder$th-order Bragg beam splitters featuring parasitic paths and open ports (dashed lines), as shown in (A) and (B) in the inertial frame of the optical lattice. Imperfections arise from population of undesired diffraction orders and affect the atom numbers in the detected ports $\portI,\portII$. Dominant parasitic paths for fifth-order of Bragg diffraction (thick dashed blue lines). Lower panels depict Gaussian pulse profiles $\Omega(t)$ equally spaced by an interrogation time $T$ and having individual pulse widths $\tau_{\mathrm{BS,M}}$. While peak values $\Opeak$ are identical in (A), a tailored choice of $\Opeak_\mathrm{M}$ enables the deflection of undesired paths and thus the suppression of the dominant parasitic interference effects in (B), see main text. In the following, we will refer to the depicted configurations as MZ types (A) and (B).
}\label{fig:MZGeometry}
\end{figure}

To date, all atom interferometers demonstrating metrological gain from LMT separations~\cite{Parker2018,Asenbaum2017,Overstreet2022} use beam splitters based on the elastic scattering of atoms from time-dependent optical lattice potentials, i.e., Bragg diffraction~\cite{Martin1988,Giltner1995PRA}. Compared to the standard picture of a two-mode interferometer, higher-order Bragg processes feature undesired diffraction orders~\cite{Mueller2008PRA,Siemss2020}, see configuration (A) in  Fig.~\ref{fig:MZGeometry}, causing systematic uncertainties on the $\mathrm{mrad}$-level referred to as diffraction phase~\cite{Buchner2003,Estey2015,Plotkin-Swing2018,Geiger2020,Narducci2022}. Yet, a comprehensive analytical model of Bragg interferometers is still missing.

%%% FIG. 1 - Panel: Trajectories, Peak Rabi frequencies
%%

In this article, we present an analytical theory for atom interferometery that takes into account the multi-port as well as multi-path physics of Bragg diffraction. Using the popular Mach-Zehnder~(MZ) geometry as an example, we present a conceptually straightforward way to suppress the diffraction phase simply by a suitable choice of pulse parameters. Indeed, a tailored combination of laser intensity and duration of the Bragg mirror pulse prevents the dominant parasitic paths from closing interferometers, as illustrated in Fig.~\ref{fig:MZGeometry}(B). By largely suppressing the parasitic interferometer paths, the physics of the diffraction phase is dramatically simplified and reduces to a phase offset that can be readily determined. This way, we reduce the remaining diffraction phase below the $\mathrm{mrad}$-level for almost the entire range of beam splitter pulse parameters that yield diffraction losses $<10\%$, compatible with efficient high-order Bragg pulses~\cite{Mueller2008PRA,Siemss2020}. We verify the accuracy of our model by comparison to simulations of the MZ interferometer in numerical experiments.

In addition to the systematic effects, the complex contributions of parasitic paths and undetected, open ports in Bragg interferometers, depicted in Fig.~\ref{fig:MZGeometry}, also affect their statistical properties. The statistical uncertainty is defined by the (Quantum) Cramér-Rao bounds~((Q)CRB)~\cite{Helstrom1969}, which have not yet been determined for Bragg interferometers, although they modify the standard-quantum limit. We calculate the (Q)CRB and show that it exhibits a nontrivial dependence on the inevitable atom loss associated with Bragg diffraction. Moreover, we demonstrate that the phase estimation strategies we present in this work allow saturation of these fundamental bounds. Our detailed study of the atomic projection noise of efficient Bragg interferometers crucially establishes important design criteria for operating these devices at or below the standard quantum limit~\cite{Salvi2018,Shankar2019,Szigeti2020,Corgier2021}.

%%%%%%
%%% Set-up: A) control parameters
%%%%%%
\textcolor{black}{\textit{Analytical model of Bragg MZ interferometer.---}} Bragg beam splitters and mirrors are generated by diffraction from an optical lattice potential $\displaystyle V(t)=2\hbar \Omega(t) \cos{(k \hat{z}-\delta t +\Lphase(t))}$. 

The lattice is pulsed by adjusting the intensity and relative frequencies, $\delta\equiv \omega_1-\omega_2$, of two counterpropagating light fields to coherently impart to the atoms a multiple of twice the photon recoil, $2\BraggOrder\hbar k$, and a lattice phase, $\Lphase(t)$. The latter is determined by the relative phase between the two  lattice laser fields. The integer $\BraggOrder$ denotes the Bragg order and the right choice of the velocity of the lattice $v_\mathrm{L}=\delta/k$ ensures resonance. Here, we focus on smooth Gaussian two-photon Rabi frequencies, $\Omega(t)=\Omega\,e^{-t^2/2\tau^2}$, reducing population of undesired diffraction orders~\cite{Mueller2008PRA}. Beam splitter and mirror operations can be realized with various combinations of peak Rabi frequency $\Omega$ and pulse width $\tau$ if they fulfill the respective condition on the pulse area~\cite{Mueller2008PRA,Siemss2020}. This freedom affords an optimized choice tailored to the experiment by balancing scattering to multiple diffraction orders against losses from velocity selectivity due to the Doppler effect~\cite{Szigeti2012,Siemss2020}, predominant for short and long pulses, respectively.

%%%%%%
%%% Set-up: B) Mach-Zehnder interferometer
%%%%%%
\begin{figure}[t]
    \includegraphics[width=1\columnwidth]{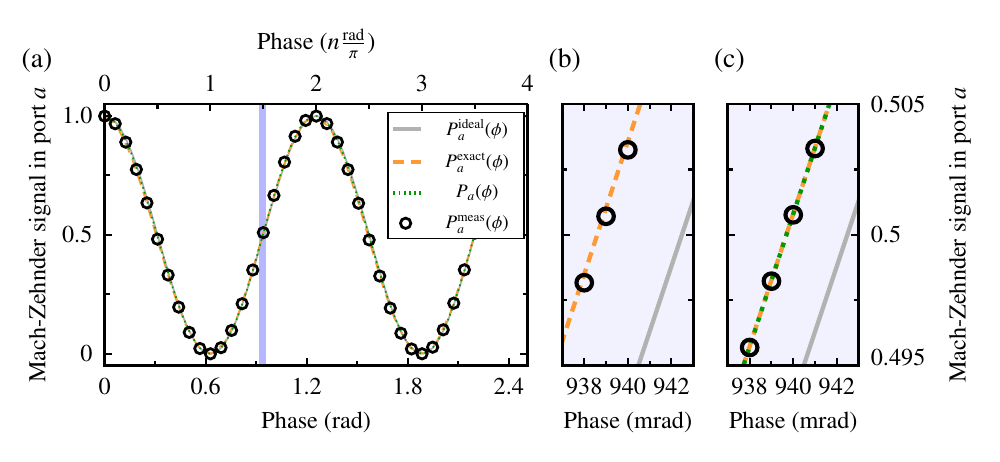}
    \caption{Signal phase scan. (a,b) Signal of MZ type (A) with $\BraggOrder=5$ in port $\portI$ scanned via the lattice phase $\Lphase(t)=\Lphase$ of the final beam splitter. We compare numerically simulated data $P^{\mathrm{meas}}_\portI(\phi)$ (symbols) to an ideal sinusoidal signal, $P^\mathrm{ideal}_\mathrm{\portI}(\phi)\equiv 1/2\cdot(1 + \cos{(\BraggOrder\phi))}$ (gray solid line), and the analytical predictions of Eqs.~\eqref{eq:GeneralEstimator} (orange dashed line) and Eq.~\eqref{eq:Estimator} (green dotted line). Pulse parameters are $\locked = 28.5\omegarec, 0.309\omegarec^{-1},0.681 \omegarec^{-1}$, with $\omegarec= \hbar k^2/2m$ being the recoil frequency of an atom with mass $m$. Beam splitter diffraction losses of  $\approx1.4\%$ amplify the signal distortions. (b) shows a bias phase shift of the ideal sinusoidal signal on the $\mathrm{mrad}$-level around mid fringe, i.e., $P_\portI(\localPhase)\approx 0.5$. In (c) mirror pulse parameters $\Mtuple =31.8\omegarec, 0.463\omegarec^{-1}$ suppress dominant interference effects as illustrated in Fig.~\ref{fig:MZGeometry}(B). This provides excellent agreement between the numerics and both analytical models in Eqs.~\eqref{eq:GeneralEstimator} and Eq.~\eqref{eq:Estimator}.}
    \label{fig:MZSignals} 
\end{figure}%
In a MZ interferometer, two identical beam splitters and a mirror pulse generate and recombine out of a wave packet several copies of different momenta. The pulses are separated by an interrogation time $T$ and will be characterized in the following by $\Opeak_\mathrm{BS},\tauBS$ and $\Opeak_\mathrm{M},\tauM$. 

Previously, we have modeled the physics of individual Bragg pulses by transfer matrices~\cite{Siemss2020}. Here, we extend this treatment to also inlcude the dominant parasitic diffraction orders, see the Supplemental Material~(SM). Transfer matrices that account for the diffraction operations and free propagation are combined to obtain the scattering matrix of a complete MZ interferometer $\mathcal{S}_\mathrm{MZ}(\phi, T,\Opeak_\mathrm{BS}, \tauBS,\Opeak_\mathrm{M},\tauM)$, which, in addition to the parameters introduced above, depends on the metrological phase $\phi$ to be measured. Furthermore, we consider an incoming atomic ensemble with average momentum $-\BraggOrder\hbar k$ relative to the optical lattice and a Gaussian momentum width well below the lattice recoil $\pspread \ll \hbar k$, consistent with recently realized, ultracold Bose-Einstein condensate sources~\cite{Kovachy2015,Deppner2021}, as initial state $\ket*{\inputState(\pspread)}$. Putting the two together, we arrive at an expression for the output state $\ket*{\outputState(\phi, T,\Opeak_\mathrm{BS}, \tauBS,\Opeak_\mathrm{M},\tauM,\pspread)} =  \mathcal{S}_\mathrm{MZ}\ket*{\inputState(\pspread)}$. Its construction and explicit form are detailed in the SM. 

From $\ket*{\outputState}$ follows the expected form of the signal for the relative atom numbers in the detected ports, 
$\displaystyle P_{\portI}(\phi) = \AtomsPortI / \left(\AtomsPortI+\AtomsPortII\right)$ and similarly for port $\portII$, cf. Fig.~\ref{fig:MZGeometry}. In atom interferometry, the measurement of relative populations $P^\mathrm{meas}_{\portI}(\phi)$ ideally suppresses the statistical fluctuations of the initial atom number $\meanAtoms$ in $\ket*{\psi^{\mathrm{in}}}$. Phase estimation usually requires scanning the phase $\phi$, e.g. via the control of the lattice phase $\Lphase$ in the experiment, in order to fit the analytic model $P_{\portI}(\phi)$ to the interferometer signal $P^\mathrm{meas}_{\portI}(\phi)$. Subsequently, the inversion $P_{\portI}(\phi)$ yields the phase estimate $\phi^\mathrm{est}=P_{\portI}^{-1}(P^\mathrm{meas}_{\portI})$. Therefore, the quality of the model $P_{\portI}(\phi)$ crucially determines the systematic accuracy and the statistical sensitivity of the phase measurement, both of which we discuss below.

\textcolor{black}{\textit{Interferometer including parasitic paths.---}}
In an interferometer realized by higher-order Bragg pulses with a generic set of parameters $\Opeak_\mathrm{BS}, \tauBS,\Opeak_\mathrm{M},\tauM$, undesired diffraction orders will populate parasitic interferometer paths and open output ports illustrated in Fig.~\ref{fig:MZGeometry}(A) to varying degrees. Their contributions to the interferometer signal can be significant and have been observed in experiments~\cite{Altin2013,Parker2016,Beguin2022}. Our analytical model for $\ket{\psi_\mathrm{out}}$ reflects this in a signal for the relative atom number  measurement, which takes the form of an infinite
Fourier series,
\begin{align} \label{eq:GeneralEstimator}
        P^{\mathrm{exact}}_{\portI}(\phi) = P_0 + \sum^{\infty}_{j=1} A_j \cos{\left(j\,\phi +\varphi_j\right)}.
\end{align}
The amplitudes $A_j$ and phases $\varphi_j$ in Eq.~\eqref{eq:GeneralEstimator} can be calculated as explained in the SM. We contrast this result with the standard model of an $\BraggOrder$-th-order Bragg MZ interferometer, which is obtained by idealizing the beam splitters and mirror as two-mode operations, $\displaystyle P^\mathrm{ideal}_{\portI(\portII)}(\phi) = P_0 \pm A \cos{\left(\BraggOrder \phi\right)}$.

In Fig.~\ref{fig:MZSignals}(a,b) we demonstrate good agreement between the analytical signal of Eq.~\eqref{eq:GeneralEstimator}, taking into account the dominant parasitic paths for $\BraggOrder=5$, and data from numerical simulations. These are based on one-dimensional descriptions of the complete matter-wave interferometer in position~space as per~\cite{Fitzek2020}, whereby all diffraction orders are fully accounted for. In Fig.~\ref{fig:MZSignals}(a,b), we scan $\phi$ via the phase of the final beam splitting pulse, which we control by selecting a value for $\Lphase = \Lphase(t)$ for each data point.  We consider an example of a set of pulses as in Fig.~\ref{fig:MZGeometry}(A) with identical peak Rabi frequency, $\Opeak=\Opeak_\mathrm{BS}=\Opeak_\mathrm{M}$, and durations that satisfy the corresponding pulse area conditions~\cite{Mueller2008PRA,Siemss2020} realizing efficient Bragg diffraction with about~$1.4\%$ losses from the main paths per the first beam splitting pulse. Even for this small population of undesired diffraction orders, the sinusoidal signal of an ideal two-mode interferometer is inadequate and exhibits a diffraction phase shift of several mrad, cf. Fig.~\ref{fig:MZSignals}(b).

Evidence of undesired additional Fourier components contributing to the signal is found in~\cite{Altin2013,Parker2016,Beguin2022}, the origin of which can be understood as follows: (i) Figure~\ref{fig:MZGeometry} makes evident that the multi-port nature of the Bragg beam splitters and the associated interference render the combined atom number in the detected ports $\portI$ and $\portII$, $\AtomsPortI+\AtomsPortII=\meanAtoms-N_{\mathrm{open}}(\phi)$, phase-dependent. Here, $N_{\mathrm{open}}(\phi)\big)$ denotes the population of all undetected (open) output ports, cf. Fig.~\ref{fig:MZGeometry}. This is in contrast to an ideal two-mode interferometer, where $\AtomsPortI+\AtomsPortII$ simply amounts to the total number of atoms $\meanAtoms$. Accordingly, the relative atom numbers in a Bragg interferometer, $\displaystyle  N_{\portI(\portII)}(\phi) / \big(N_{\mathrm{atom}}-N_{\mathrm{open}}(\phi)\big)$, will be a ratio of $\phi$-dependent functions and therefore in general contain Fourier components of any order. 
(ii) In addition, the functional dependence of the absolute atom numbers $N_{\portI,\portII}(\phi)$ is also complicated by the occurrence of parasitic interferometers. For the example of the MZ interferometer in Fig.~\ref{fig:MZGeometry}(A) parasitic MZ terms (cf.~\cite{Altin2013}) arise, when undesired diffraction orders overlap with the main interferometry arms at $t=2T$. In particular, this causes $A_j$ and $\varphi_j$ to depend on the interrogation time $T$, in addition to the parameters of the individual Bragg pulses, as observed in~\cite{Altin2013,Parker2016}. Notwithstanding its correctness, it will be challenging to apply the waveform in Eq.~\eqref{eq:GeneralEstimator} due to the large number of parameters involved and the limited control over them.

\textcolor{black}{\textit{Interferometer with suppressed parasitic paths.---}} A simple way to efficiently suppress interference with the dominant parasitic paths in the MZ geometry is to design the mirror pulse to be approximately transparent to them, as illustrated in Fig.~\ref{fig:MZGeometry}(B). Specific combinations $\Mtuple$ achieving this for $\BraggOrder=5$ are stated in the SM but they do in fact exist for all relevant higher orders of Bragg diffraction, except $\BraggOrder=2$. It is straightforward to omit the influence of parasitic paths in our analytical model and consider only the effects of the open ports (point (i) above). Doing so simplifies the signal of the MZ interferometer in ports $\portI$ and $\portII$ to
\begin{align} \label{eq:Estimator}
\begin{split}
    P_{\portI(\portII)}(\phi) &= P_0 \pm \sum^{3}_{j=1} A_j \sin{\left(j\cdot \left(\BraggOrder\phi+\gamma + \frac{\pi}{2}\right)\right)} + \mathcal{O}[\gamma^3], 
\end{split}
\end{align}
where $P_{\portII}(\phi)$ is shifted by $\pi$ relative to $P_{\portI}(\phi)$ as one would expect, see SM. In contrast to Eq.~\eqref{eq:GeneralEstimator} this expression contains only the harmonics of a single Fourier component $\BraggOrder\phi$, and a phase shift $\gamma$ common to all harmonics. Moreover, we can show that this shift is independent of $T$ and can be calculated for given beam splitter parameters $\BStuple$, see SM. In fact, $\gamma$ is a small parameter closely related to the losses to undesired diffraction orders during beam splitting~\cite{Siemss2020}. For an interferometer with suppressed parasitic paths as in Fig.~\ref{fig:MZGeometry}(B), both the exact signal in Eq.~\eqref{eq:GeneralEstimator} and the much simpler formula in Eq.~\eqref{eq:Estimator} are in excellent agreement with the data from a numerical simulation, as can be seen in Fig.~\ref{fig:MZSignals}(c).

\begin{figure}[t]
        \includegraphics[width=0.45\textwidth]{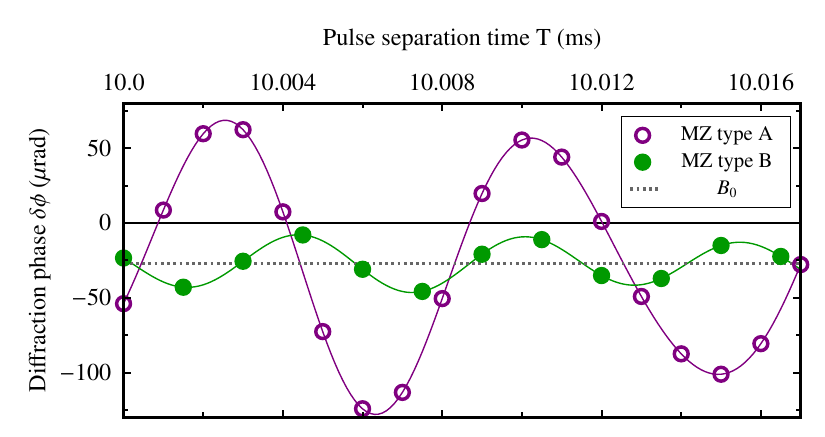}
        \caption{ Phase estimation error. We evaluate the diffraction phase $\diffPhase$ in Eq.~\eqref{eq:PhaseOffset} using numerical simulations of MZ interferometers of types (A) and (B) with $\BraggOrder=5$. $\diffPhase$ is determined at mid fringe $\localPhase = \frac{3\pi}{2} \frac{1}{5}$ and beam splitters are chosen to operate at a so-called 'magic' Bragg duration minimizing individual diffraction losses to $\approx0.18\%$ (see main text). This reduces oscillations in $\diffPhase$ to the $\mu\mathrm{rad}$-level while scanning the pulse separation time $T$ (cf.~\cite{Altin2013,Parker2016}). The first data set (open circles) has been obtained for parameters $\locked = 30.75\omegarec, 0.218\omegarec^{-1},0.519 \omegarec^{-1}$, with $\omegarec= \hbar k^2/2m$ being the recoil frequency of an atom with mass $m$.  Suppressing dominant parasitic interference effects via $\Mtuple =31.8\omegarec, 0.463\omegarec^{-1}$ (closed circles, case (B)) further reduces the oscillation amplitude by a factor of five to $ <40\,\mu\mathrm{rad}$. Solid lines represent fits to the data explained in the main text, the offset of which is the effectively same $B_0\approx - 27\, \mu\mathrm{rad}$ (dotted) for both data sets.
        }\label{fig:DiffPhaseLZMin}
\end{figure}%

\textcolor{black}{\textit{Diffraction phase.---}}
We proceed to quantify the diffraction phase~\cite{Estey2015,Parker2016}, i.e., the systematic deviation of the model in Eq. ~\eqref{eq:Estimator} from the actual signal obtained in numerical simulations. We compare its application to the signal of interferometers~(A) without and (B) with suppression of parasitic paths, as in Fig.~\ref{fig:MZGeometry}(A) and (B) respectively. The deviation between the phase and its estimate $\phi^\mathrm{est}=P_{\portI}^{-1}(P_a^\mathrm{meas})$ is
\begin{align}\label{eq:PhaseOffset}
        \diffPhase = P_{\portI}^{-1}(P_a^\mathrm{meas}) - \phi=\left.P_{\portI}^{-1}(P_a^\mathrm{meas})\right|_{\gamma=0} - \phi -\frac{\gamma}{\BraggOrder}.
\end{align}
Here, we emphasize the fact, that $\gamma$ is a shift common to all Fourier components in Eq.~\eqref{eq:Estimator}, which we have taken advantage of in the second equality. The error $\delta\phi$ in determining the actual phase $\phi$ depends on the knowledge of the value of $\gamma$ and the accuracy of the model with respect to the remaining phase, $\left.P_{\portI}^{-1}(P_a^\mathrm{meas})\right|_{\gamma=0}$. Since $\gamma$ can be inferred quite accurately for given beam splitter parameters (cf. SM) it is the latter contribution which sets the systematic uncertainty. Furthermore, the scaling $\BraggOrder \phi$ in Eq.~\eqref{eq:Estimator} highlights, that both contributions to the diffraction phase in Eq.~\eqref{eq:PhaseOffset} will be linearly suppressed by the order $\BraggOrder$ of the Bragg pulses, cf.~\cite{Estey2015}.

We extract $\diffPhase$ after fitting the analytical model Eq.~\eqref{eq:Estimator} to signals $P_a^\mathrm{meas}$ generated in  numerical simulations and plot it in Fig.~\ref{fig:DiffPhaseLZMin} against the pulse separation time $T$ for interferometers of type (A) and (B). 
In both realizations of the MZ geometry with fifth-order Bragg pulses, spurious MZ interference terms cause oscillations at frequencies $(5\pm 1)\omegarec$, which are related to the recoil frequency $\omegarec= \hbar k^2/2m$ via the kinematic phases of the main interferometer arms relative to the main parasitic diffraction orders, cf.~\cite{Altin2013,Parker2016}. Here, $m$ is the mass of the atom. This is despite minimizing beam splitter losses to about $0.18\%$ via our selection of pulse parameters, referred to by \textit{Parker et al.} as a 'magic' Bragg duration, which effectively reduces the $T$-dependence of the diffraction phase in the $\BraggOrder=5$ conjugated Ramsey-Bordé interferometer discussed in Ref.~\cite{Parker2016}. We can use a fit model $f(T) = B_0 + B_1 \cos{\left(4\omegarec T + \eta_1\right)}+ B_2 \cos{\left(6\omegarec T + \eta_2\right)}$ to extract the amplitude offset $B_0$ and the peak-to-peak value $\mathrm{PP}\equiv \abs{ \max\nolimits_{\forall T } f(T) -  \min\nolimits_{\forall T } f(T)}$ of the oscillations in the diffraction phase. For the pulse parameters assumed in Fig.~\ref{fig:DiffPhaseLZMin}, the offset $B_0\approx -27 \, \mu\mathrm{rad}$ is the same for both cases (A) without and (B) with suppression of parasitic paths. This shows, that the inclusion of $\gamma/\BraggOrder\approx 280\,\mu\mathrm{rad}$ in Eq.~\eqref{eq:PhaseOffset} accounts for most of the $T$-independent shift. However, $\mathrm{PP}$ values of both data sets are very different, lying in the range of 200\,$\mu$rad for (A) and about 40\,$\mu$rad for (B). This reduction is significant because the net diffraction phase shift may be on the order of the PP value due to insufficient control over the separation time $T$ at the $\mu$s level or, if $T$ is sampled, due to aliasing effects, cf.~\cite{Parker2016}.

\begin{figure}[ht]
        \includegraphics[width=0.48\textwidth]{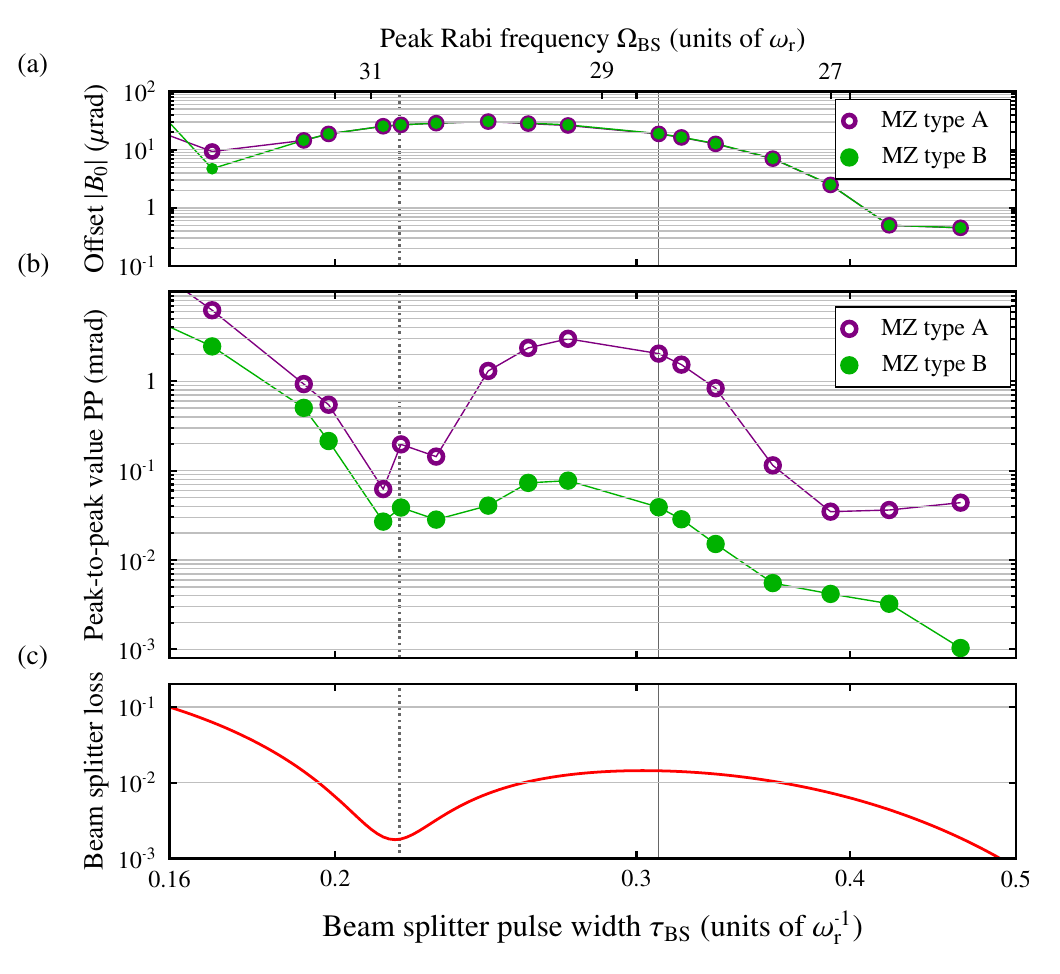}
        \caption{
       Diffraction phase suppression in MZ geometry. (a,b) Fits to Eq.~\eqref{eq:PhaseOffset} are evaluated for numerical MZ signals recorded over scans of $T\in [10,10.016]~\mathrm{ms}$ as shown in Fig.~\ref{fig:DiffPhaseLZMin}(a) using pulse parameter respecting the condition on the Bragg pulse area~\cite{Mueller2008PRA,Siemss2020}. Closed (open) symbols represent data with (w/o) suppression of parasitic paths parameterized via $\Opeak = \Opeak_\mathrm{BS},\tauBS$. (a) shows absolute values of the residual offset $|B_0|\leq30\,\mu\mathrm{rad}$ (symbols) independent of the mirror pulse.  (b) The adapted mirror pulse (case (B)) significantly suppresses PP values for most parameters below $1~\mathrm{mrad}$ and to less than $10~\mu\mathrm{rad}$ for sufficiently long beam splitter pulse durations. Lines connecting the data points serve as guides to the eye.
       (c) Numerically obtained population loss from the main diffraction orders $\ket{\pm 5\hbar k}$ of a single beam splitter. Results in Fig.~\ref{fig:MZSignals} (Fig.~\ref{fig:DiffPhaseLZMin}) use parameters corresponding to the visible local maximum (minimum) denoted by the solid (dotted) vertical line.}
        \label{fig:DiffPhaseSuppression}
\end{figure}

The offset $|B_0|$ and the $\mathrm{PP}$ value of the oscillations in the diffraction phase are given in Fig.~\ref{fig:DiffPhaseSuppression}(a,b) for a range of parameters $\BStuple$, which we restrict by requiring beam splitter losses of less than $10\%$. We again compare both Bragg mirror pulse configurations (A) and (B). Figure~\ref{fig:DiffPhaseSuppression}(a) confirms that through the inclusion of $\gamma$ in Eq.~\eqref{eq:Estimator} we achieve a residual $T$-independent contribution to the diffraction phase $\diffPhase$ of at most a few tenths of $\mu\mathrm{rad}$ for both Bragg mirror configurations. At the same time, Fig.~\ref{fig:DiffPhaseSuppression}(b) highlights that the oscillations of $\diffPhase$ can be on the order of several $\mathrm{mrad}$ and are therefore comparable to $\gamma /n$, see SM, in case of  pulse parameters with relatively strong couplings to undesired diffraction orders. As implied by Fig.~\ref{fig:DiffPhaseSuppression}(c), the behavior of both quantities characterizing $\diffPhase$ is directly related to the losses in beam splitter operations. Notably, the local minimum indicates the magic Bragg duration for fifth-order Bragg beam splitters mentioned earlier but in fact, such minima exist for all orders $\BraggOrder>1$ and are a feature predicted by Landau-Zener theory~\cite{Siemss2020}.

We conclude that the diffraction phase of MZ Bragg interferometers can be suppressed by means of the Bragg mirror pulse below $1~\mathrm{mrad}$ for most parameters $\BStuple$ and even down to a few $\mu\mathrm{rad}$ for sufficiently long beam splitter durations. In contrast, without suppression of parasitic paths and without account for the diffraction phases (including $\gamma$) by means of Eq.~\eqref{eq:Estimator}, the accuracy would be limited to more than $0.5~\mathrm{mrad}$ in the same regime, which constitutes an improvement by two orders of magnitude. The remaining diffraction phase is limited by higher-order contributions in $\gamma$ and the finite efficiency in suppressing parasitic paths.
%%%%%%
%%%  Result: C.3) Adapted Mirror 
%%%%%%

%
%%
%%% FIG. - Panel: Phase estimation offset vs pulse parameters for \delta p =0.01hk %%
%

%%%%%%
%%%  Phase Sensitivity 
%%%%%%
\begin{figure}[ht]
    \includegraphics[width=0.47\textwidth]{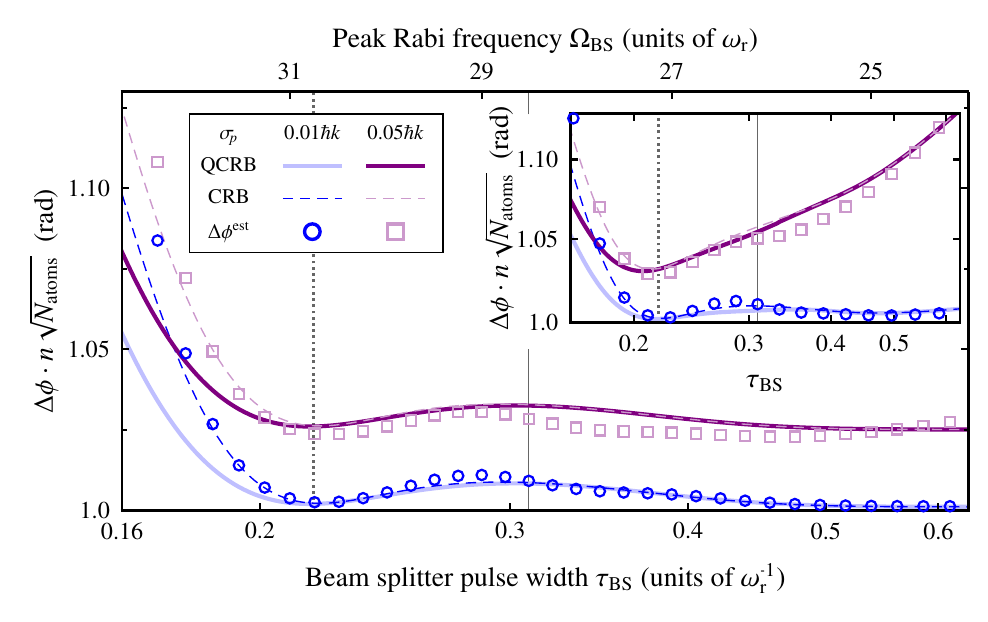}
    \caption{Sensitivity bound of multi-port Bragg interferometers with uncorrelated particles. Cramér-Rao bound (CRB, dashed lines) and Quantum Cramér-Rao bound (QCRB, solid lines) for a fifth-order Bragg MZ interferometer in configuration (B). Upon scanning $\BStuple$ we suppress the dominant parasitic interferometery paths using pulse parameters $\Mtuple =31.8\omegarec, 0.463\omegarec^{-1}$. Both bounds are shown for momentum spread $\pspread=0.01\hbar k~(0.05\hbar k)$ in blue (purple) and scaled to the projection noise limit of an ideal two-mode MZ, $\BraggOrder\sqrt{N_\mathrm{atoms}}$. Symbols represent the statistical uncertainty $\statUncert$ of a phase estimate based on Eq.~\eqref{eq:Estimator} at mid fringe obtained for numerical simulations. We set $\localPhase = \frac{3\pi}{2} \frac{1}{5}$ for both the CRB and for $\statUncert$ and fix the time $T=10~\mathrm{ms}$. The statistical uncertainty  operated with equal Peak Rabi frequencies (case (A) in inset) suffers from increased finite-velocity Doppler effects.}\label{fig:StatUncertainty}
\end{figure}

\textcolor{black}{\textit{Phase sensitivity.---}}
To complete our study of the model in Eq.~\eqref{eq:Estimator}, we discuss the statistical uncertainty of the phase estimate $\phi^\mathrm{est}=P_{\portI}^{-1}(P_a^\mathrm{meas})$, which is given by the projection noise $ \statUncert \approx   \sqrt{\frac{P_\portI(\phi) (1- P_\portI(\phi))}{\AtomsPortI+\AtomsPortII}} \frac{1}{\abs{\partial_{\phi} P_\portI(\phi) }} $, cf. SM. We emphasize that this is quite different from the projection noise of an ideal two-mode MZ-Bragg interferometer, which is $\statUncert = \frac{\sqrt{P_0(1-P_0)}}{\BraggOrder\, A \sqrt{\meanAtoms}}$ at mid fringe, where again $\meanAtoms$ denotes the number of atoms entering the interferometer. We benchmark the achievable phase sensitivity by the CRB and the QCRB, both of which follow from the analytical expression for $\ket*{\outputState}$, cf. SM. The CRB sets the projection noise limit when measuring relative atom numbers in the main ports $\portI,\portII$, as depicted in Fig.~\ref{fig:MZGeometry}. The QCRB bounds the projection noise for arbitrary measurements performed on all output ports of the final Bragg beam splitter. 

The results are shown in Fig.~\ref{fig:StatUncertainty} for MZ interferometers of type (B) and, in the inset, for (A) for the same range of beam splitter parameters considered in Fig.~\ref{fig:DiffPhaseSuppression}. First of all, we observe that $\statUncert$ for phase estimation based on Eq.~\eqref{eq:Estimator} using numerical data agrees well with the analytical  CRB and QCRB. The visible deviations are on the level to be expected due to our perturbative treatment of finite velocity effects and beam splitter losses, cf. SM. We show the phase sensitivity in Fig.~\ref{fig:StatUncertainty} scaled to $\BraggOrder\sqrt{N_\mathrm{atoms}}$, i.e., the CRB of an ideal two-mode interferometer with $P_0=A=1/2$. This reveals that the projection noise limit for a Bragg interferometer lies a few percent above this value. The increase of the CRB with growing momentum spread $\pspread$ is caused by atom losses due to velocity selectivity of the Bragg process, which become stronger for longer pulse durations, cf.~\cite{Szigeti2012}. In addition, as revealed by our two choices of velocity width, velocity selectivity is reduced for case (B) because of the relatively short mirror pulse duration for the suppression of parasitic paths, see SM. The loss of sensitivity at shorter beam splitter pulse durations is due to the increasing non-adiabaticity of Bragg diffraction and the associated diffraction losses~\cite{Siemss2020}, see Fig.~\ref{fig:DiffPhaseSuppression}(c). Interestingly, there is no discernible difference in performance between either configuration despite deliberately deflecting atoms out of the interferometer in scenario (B), see Fig.~\ref{fig:MZGeometry}(B). The reason being, that Bragg diffraction losses primarily populate parasitic interferometers with scale factors smaller than the main diffraction order (here, $\BraggOrder=5$). Accordingly, their contributions effectively decrease the space-time area and thus increase the statistical uncertainty of the phase measurement. Overall, best sensitivity is achieved at the local minimum of beam splitting losses from the main diffraction orders $\pm 5\hbar k$, cf. Fig.~\ref{fig:DiffPhaseSuppression}(c). This sets the fundamental projection noise limit of a Bragg atom interferometer.

%%%%%%%%%%%%%%%%%%%%%%%%%%%%%%%%%%%%%%%%%%%%%%%%%%%%%%%%%%%%%%%%
%%% Conclusions
%%%%%%%%%%%%%%%%%%%%%%%%%%%%%%%%%%%%%%%%%%%%%%%%%%%%%%%%%%%%%%%% 
\textcolor{black}{\textit{Conclusions.---}} In summary, we have presented an analytical model for LMT atom interferometers based on Bragg diffraction, which permits a thorough understanding of their systematic and statistical uncertainties and their fundamental sensitivity bounds. Our model provides design criteria for reaching these bounds and paves the way towards accuracies in the $\mu$rad-range using higher-order Bragg diffraction in combination with ultra-cold atomic sources~\cite{Kovachy2015,Deppner2021}. The operation of LMT interferometers at or near the limit of quantum projection noise is a critical requirement if they are to be combined with entangled sources~\cite{Szigeti2020,Corgier2021}. The methods and techniques developed here are general and can be applied also to other interferometer topologies, such as the conjugated Ramsey-Bordé interferometer in~\cite{Parker2018}, and to other diffraction techniques as, e.g., double Bragg diffraction pulses~\cite{Giese13PRA,Ahlers2016,Jenewein2022}. Our work contributes to the development of high-precision quantum sensors for fundamental tests and towards atom interferometers fulfilling the size, weight, and power (SWaP) requirements of modern real-world applications~\cite{Geiger2020,Narducci2022}, especially in combination with resonator-enhanced light fields~\cite{Hamilton2015}.

%%%%%%%%%%%%%%%%%%%%%%%%%%%%%%%%%%%%%%%%%%%%%%%%%%%%%%%%%%%%%%%%
%%%     Section: Acknowledgements
%%
%%%%%%%%%%%%%%%%%%%%%%%%%%%%%%%%%%%%%%%%%%%%%%%%%%%%%%%%%%%%%%%% 
\begin{acknowledgements}
 We thank A. Gauguet and S. Loriani for helpful comments on the manuscript. This work was funded by the Deutsche Forschungsgemeinschaft (German Research Foundation) under Germany’s Excellence Strategy (EXC-2123 QuantumFrontiers Grant No. 390837967), through CRC 1227 (DQ-mat) within Projects No. A05, No. B07 as well as No. B09 and QuantERA project 499225223 (SQUEIS), the Verein Deutscher Ingenieure (VDI) with funds provided by the German Federal Ministry of Education and Research (BMBF) under Grant No. VDI 13N14838 (TAIOL), and the Ger\-man Space Agency (DLR) with funds provided by the German Federal Ministry of Economic Affairs and Energy (BMWi) due to an enactment of the German Bundestag under Grant No. DLR 50WM1952 (QUANTUS-V-Fallturm), 50WP1700 (BECCAL), 50WM2245A (CAL-II), 50WM2060 (CARIOQA), 50WM2253A ((AI)²) as well as 50RK1957 (QGYRO). We furthermore acknowledge financial support from "Niedersächsisches Vorab" through "Förderung von Wissenschaft und Technik in Forschung und Lehre" for the initial funding of research in the new DLR-SI Institute and  the “Quantum- and Nano Metrology (QUANOMET)” initiative within the project QT3.
\end{acknowledgements}

%%%%%%%%%%%%%%%%%%%%%%%%%%%%%%%%%%%%%%%%%%%%%%%%%%%%%%%%%%%%%%%%
%%%     Bibliography
%%%%%%%%%%%%%%%%%%%%%%%%%%%%%%%%%%%%%%%%%%%%%%%%%%%%%%%%%%%%%%%% 
\bibliographystyle{apsrev4-1_custom}
\bibliography{bibliography}

%%%%%%%%%%%%%%%%%%%%%%%%%%%%%%%%%%%%%%%%%%%%%%%%%%%%%%%%%%%%%%%% End of document
%%%%%%%%%%%%%%%%%%%%%%%%%%%%%%%%%%%%%%%%%%%%%%%%%%%%%%%%%%%
\end{document}

\title{Supplemental material: Large-momentum-transfer atom interferometers with \texorpdfstring{$\mu$}{u}rad-accuracy using Bragg diffraction}

\def\affitp  {\affiliation{Leibniz Universit\"at Hannover, Institut f\"ur Theoretische Physik, Appelstr. 2, D-30167 Hannover, Germany }}
\def\affiqo  {\affiliation{Leibniz Universit\"at Hannover, Institut f\"ur Quantenoptik, Welfengarten 1, D-30167 Hannover, Germany }}
\def\affdlr  {\affiliation{Deutsches Zentrum f\"ur Luft- und Raumfahrt e. V. (DLR),\\
Institut f\"ur Satellitengeod\"asie und Inertialsensorik, Callinstraße 30b, D-30167 Hannover, Germany}}
\author{J.-N.~Siem\ss}
\email[email (he/him/his): ]{jan-niclas.siemss@itp.uni-hannover.de}
\affitp
\affiqo
\author{F.~Fitzek}
\affitp
\affiqo
\author{C.~Schubert}
\affiqo
\affdlr 
\author{E.~M.~Rasel}
\author{N.~Gaaloul}
\affiqo
\author{K.~Hammerer}
\email[email: ]{klemens.hammerer@itp.uni-hannover.de}
 \affitp
\date{\today}

%\let\clearpage\relax

% \begin{abstract}
% \end{abstract}

\maketitle

\tableofcontents
\section{Scattering matrix formalism for Bragg diffraction on pulsed optical lattices}\label{App:FullScatteringMatrix}
In this section of the supplemental material, we construct the scattering matrix $\mathcal{S}_\mathrm{MZ}(\phi, T,\BStuple,\Mtuple)$ describing the atom-light interaction in the case of the symmetric Mach-Zehnder~(MZ) interferometer discussed in the main article.
%%%%%%%%%%%%%% 
%%%%%% SubSection: Scattering Matrix for individual Bragg pulses
%%%%%%%%%%%%%% 
\subsection{Scattering matrix for individual Bragg pulses}\label{App:BraggScatteringMatrix}
The transfer matrices $S$ for single Bragg operations have been derived in~\cite{Siemss2020} accounting for the main diffraction orders. Here, we extend this treatment in order to include effects of parasitic diffraction orders and parasitic interferometer paths.
We adopt the notation of~\cite{Siemss2020}, which uses a lattice of momentum states $\ket{j\hbar k+p}$ with $j\in\mathds{Z}$ and (quasi)momentum $p \in \left[-\hbar k/2,\hbar k/2 \right]$. To realize Bragg diffraction of a particular order $\BraggOrder$, the relative velocity between the atomic ensemble and the optical lattice is chosen so that efficient transitions occur between the states $\{\ket{\pm \BraggOrder\hbar k+p}\}$. The (quasi)momentum $p$ in our model accounts for the finite velocity width of the atomic cloud, enabling the inclusion of Doppler effects. In order to cover the dominant parasitic diffraction orders,  we enlarge the subspace of momentum states relevant for Bragg diffraction of order $\BraggOrder\geq 3$ to include also $\{\ket{\pm \hbar k(\BraggOrder-2)+p}\}$. Thus, for a given order of Bragg diffraction $\BraggOrder$, we will be dealing with a four-dimensional state space containing states
\begin{align}\label{SMeq:statespace}
    \ket*{q_j(p)}\in \qty\Big{\ket{\pm \BraggOrder\hbar k+p},\ket{\pm \hbar k(\BraggOrder-2)+p}}.
\end{align}
In the following, we will mostly suppress the explicit dependence of $\ket*{q_j}$ on the (quasi-)momentum $p$ for the sake of clarity and reintroduce it when appropriate. The two cases of lower order Bragg diffraction, $\BraggOrder=1,2$, can be treated analogously with an adjusted state space $\qty\big{\ket{\pm \BraggOrder\hbar k+p},\ket{\pm \hbar k(\BraggOrder+2)+p}}$  and $\qty\big{\ket{\pm \BraggOrder\hbar k+p},\ket{p}}$, respectively. However, in the following we will focus on the discussion of higher order Bragg diffraction, $\BraggOrder\geq 3$, with the state space~\eqref{SMeq:statespace}.

To derive a scattering matrix for the MZ geometry discussed in the main text, we must combine individual transfer matrices of the form
\begin{subequations}\label{SMeq:BraggMatrix_full}
	\begin{align} \label{SMeq:BraggMatrix_a}
		S(\Opeak,\tspread)=\int\displaylimits_{-\hbar k/2}^{\hbar k/2}\!\!\difp\sum_{j,l} \left[B(p,\Opeak,\tspread)\right]_{jl}
		\ketbra*{q_j(p)}{q_l(p)},
	\end{align}
	where
	\begin{align}\label{SMeq:BraggMatrix_b}
	\begin{split}
    	B(p,\Opeak,\tspread)
		=
			%\begin{pmatrix} 
			\left(\begin{array}{@{}c|c c|c@{}}
			B_{-\BraggOrder,-\BraggOrder}       & \textcolor{blue}{B_{-\BraggOrder,-(\BraggOrder-2)} }   &    \textcolor{blue}{B_{-\BraggOrder,+(\BraggOrder-2)}}& B_{-\BraggOrder,+\BraggOrder}     \BstrutTwo\\\hline
						\TstrutTwo \textcolor{blue}{B_{-(\BraggOrder-2),-\BraggOrder} }  & \textcolor{blue}{B_{-(\BraggOrder-2),-(\BraggOrder-2)}} & \textcolor{blue}{B_{-(\BraggOrder-2),+(\BraggOrder-2)}} & \textcolor{blue}{B_{-(\BraggOrder-2),+\BraggOrder}}\BstrutTwo \\
							\TstrutTwo \textcolor{blue}{B_{+(\BraggOrder-2),-\BraggOrder}}	   & \textcolor{blue}{B_{+(\BraggOrder-2),-(\BraggOrder-2)}} & \textcolor{blue}{ B_{+(\BraggOrder-2),+(\BraggOrder-2)}} & \textcolor{blue}{B_{+(\BraggOrder-2),+\BraggOrder} } \BstrutTwo\\\hline
								\TstrutTwo B_{+\BraggOrder,-\BraggOrder}       & \textcolor{blue}{B_{+\BraggOrder,-(\BraggOrder-2)}}      & \textcolor{blue}{B_{+\BraggOrder,+(\BraggOrder-2)} }    & B_{+\BraggOrder,+\BraggOrder} 
			\end{array}\right).
\end{split}
\end{align}
\end{subequations}
We indicate the sub-blocks of the transfer matrices referring to the main diffraction orders (denoted by black symbols) and parasitic orders (blue symbols). The matrix elements $B_{j,l}$ are functions of the pulse parameters $\big\{\Omega, \tau\big\}$ and the (quasi)momentum $p$. As a reference, we give the form of Eq.~\eqref{SMeq:BraggMatrix_b} in the idealized two-mode case,
\begin{align}  \label{SMeq:TwoModeBraggMatrices}
B_\mathrm{BS}^{\mathrm{ideal}} &=  \left(\begin{array}{@{}c|c|c|c@{}}
%%Row#1
1/\sqrt{2}  &\multicolumn{2}{c|}{ \textcolor{blue}{\textbf{0}} } & - i e^{-i2\BraggOrder \Lphase} / \sqrt{2}\BstrutTwo \\ \hline
%%Row#2
\multirow{2}{*}{ \textcolor{blue}{\textbf{0}}}&\multicolumn{2}{c|}{\multirow{2}{*}{\scalebox{1.3}
        {\textcolor{blue}{$\textbf{B}_{\pm(\BraggOrder-2)}$}}}}& \multirow{2}{*}{ \textcolor{blue}{\textbf{0}}}\\
%%Row#3
 &\multicolumn{2}{c|}{\multirow{2}{*}{}}& \BstrutTwo\\\hline
%%Row#4
\TstrutTwo - i e^{+i2\BraggOrder \Lphase} /\sqrt{2}&\multicolumn{2}{c|}{ \textcolor{blue}{\textbf{0}}} &  1/\sqrt{2} \BstrutTwo \\
\end{array}\right), &   B_\mathrm{M}^{\mathrm{ideal}} &= 
\left(\begin{array}{@{}c|c|c|c@{}}
%%Row#1
0 &\multicolumn{2}{c|}{ \textcolor{blue}{\textbf{0}}} & - i e^{-i2\BraggOrder \Lphase} \BstrutTwo \\ \hline
%%Row#2
\TstrutTwo \multirow{2}{*}{ \textcolor{blue}{\textbf{0}}}&\multicolumn{2}{c|}{\multirow{2}{*}{\scalebox{1.3}
        {\textcolor{blue}{$\textbf{B}_{\pm(\BraggOrder-2)}$}}}}& \multirow{2}{*}{ \textcolor{blue}{\textbf{0}}} \BstrutTwo\\
%%Row#3
\TstrutTwo  &\multicolumn{2}{c|}{\multirow{2}{*}{}}& \BstrutTwo\\\hline
%%Row#4
\TstrutTwo  - i e^{+i2\BraggOrder \Lphase} &\multicolumn{2}{c|}{ \textcolor{blue}{\textbf{0}}} & 0 \\
        \end{array}\right), 
\end{align}
describing, respectively, perfect beam splitter and mirror operation for a wave packet initially prepared with mean momentum $-\BraggOrder\hbar k $ in the limit of a vanishing velocity width. Under this assumption, parasitic diffraction orders decouple from the main ones.

Beyond this idealized case, we model the functional dependence of the scattering matrix $B$ on the pulse parameters via the application of the adiabatic theorem as in~\cite{Siemss2020}. Extending this description to the dominant spurious diffraction orders, we find the analytical form of the transfer matrix,
\begin{subequations} \label{SMeq:BMatrix}
\begin{align}
B(p,\Opeak,\tspread) &= \mathcal{M} (\Opeak,\tspread) \cdot \mathcal{N}(p,\Opeak,\tspread)
\end{align}
where
\begin{align} \label{SMeq:BMatrixpZerothOrd}
\resizebox{17.5cm}{!}{ %{16.8cm}
 \begin{math}
        \begin{aligned}
\mathcal{M}(\Omega,\tau)   &= 
\left(\begin{array}{@{}c|c c|c@{}}
%%Row #1
%[1,1]
\scalebox{\MatrixScaleSmall}{$
e^{\frac{-i}{2}\left( \glP_\BraggOrder-i\sumLoss\right)}\cos{\left(\frac{\dPhi_\BraggOrder-i\diffLoss}{2}\right)}
$}%checked       
& %[1,2]
\scalebox{\MatrixScale}{$
\begin{smallmatrix}
\textcolor{blue}{
\frac{-1}{2} e^{\frac{-i}{2} \left(\dPhi_{\BraggOrder-2} +\glP_{\BraggOrder-2} + 4 \Lphase \right)}\times}\\\textcolor{blue}{ \big( e^{i \LZphiplus}  h_1(\sumLoss,\diffLoss)+e^{i \left(\dPhi_{\BraggOrder-2} +\LZphimin\right)} h_2(\sumLoss,\diffLoss)  \big) }
\end{smallmatrix}
$}%checked
& %[1,3]
\scalebox{\MatrixScale}{$
\begin{smallmatrix} 
\textcolor{blue}{
\frac{-1}{2} e^{\frac{-i}{2} \left(\dPhi_{\BraggOrder-2} +\glP_{\BraggOrder-2} + 4(\BraggOrder-1) \Lphase \right)} \times
} \\
\textcolor{blue}{
\big(e^{i \LZphiplus} h_1(\sumLoss,\diffLoss)- e^{i \left(\dPhi_{\BraggOrder-2} +\LZphimin\right)} h_2(\sumLoss,\diffLoss) \big)  
}
\end{smallmatrix}
$}%checked  
& %[1,4]
\scalebox{\MatrixScaleSmall}{$
- i e^{-i2\BraggOrder \Lphase} e^{\frac{-i}{2}\left( \glP_\BraggOrder-i\sumLoss\right)}\sin{\left(\frac{\dPhi_{\BraggOrder}-i\diffLoss}{2}\right)} 
$}%checked
\Bstrut\\\hline
%%Row #2
%[2,1]
\Tstrut
\scalebox{\MatrixScale}{$
\begin{smallmatrix}
 \textcolor{blue}{
 \frac{1}{2} e^{\frac{-i}{2} \left(\dPhi_{\BraggOrder} +2\left(\LZphimin+\LZphiplus\right) +\glP_{\BraggOrder} - 4 \Lphase \right)} \times
 }
 \\ \textcolor{blue}{\big(e^{i \LZphimin} h_1(\sumLoss,\diffLoss) + e^{i \left(\dPhi_{\BraggOrder} +\LZphiplus\right)} h_2(\sumLoss,\diffLoss) \big)   }
\end{smallmatrix} 
$}%checked
& %[2,2]
\scalebox{\MatrixScaleSmall}{$
\textcolor{blue}{e^{\frac{-i}{2}\left( \glP_{\BraggOrder-2}-i\sumLoss\right)}\cos{\left(\frac{\dPhi_{\BraggOrder-2}-i\diffLoss}{2}\right)} } 
$}%checked 
&%[2,3] 
\scalebox{\MatrixScale}{$
\begin{smallmatrix}
 \textcolor{blue}{- i e^{-i2(\BraggOrder-2) \Lphase} e^{\frac{-i}{2}\left( \glP_{\BraggOrder-2}-i\sumLoss\right)}\sin{\left(\frac{\dPhi_{\BraggOrder-2}-i\diffLoss}{2}\right)}} 
\end{smallmatrix} 
$}%checked
& %[2,4]
\scalebox{\MatrixScale}{$
\begin{smallmatrix}
 \textcolor{blue}{
  \frac{1}{2} e^{\frac{-i}{2} \left(\dPhi_{\BraggOrder} +2\left(\LZphimin+\LZphiplus\right) +\glP_{\BraggOrder} + 4(\BraggOrder-1) \Lphase \right)} \times }
\\ \textcolor{blue}{\big( e^{i \LZphimin} h_1(\sumLoss,\diffLoss)  - e^{i\left( \dPhi_{\BraggOrder} +\LZphiplus\right)} h_2(\sumLoss,\diffLoss) \big)}
\end{smallmatrix}
$}%checked
 \Bstrut\\
%%Row #3
%[3,1]
\Tstrut
\scalebox{\MatrixScale}{$
\begin{smallmatrix}
 \textcolor{blue}{
  \frac{1}{2} e^{\frac{-i}{2} \left(\dPhi_{\BraggOrder} +2\left(\LZphimin+\LZphiplus\right) +\glP_{\BraggOrder} - 4(\BraggOrder-1) \Lphase \right)} \times }
\\ \textcolor{blue}{\big( e^{i \LZphimin} h_1(\sumLoss,\diffLoss)  - e^{i\left( \dPhi_{\BraggOrder} +\LZphiplus\right)} h_2(\sumLoss,\diffLoss) \big)} %checked
\end{smallmatrix}
$}
&%[3,2]
\scalebox{\MatrixScale}{$
\begin{smallmatrix}
 \textcolor{blue}{- i e^{i2(\BraggOrder-2) \Lphase} e^{\frac{-i}{2}\left( \glP_{\BraggOrder-2}-i\sumLoss\right)}\sin{\left(\frac{\dPhi_{\BraggOrder-2}-i\diffLoss}{2}\right)}} 
 \end{smallmatrix}
$}%checked
&%[3,3]
\scalebox{\MatrixScaleSmall}{$
\textcolor{blue}{e^{\frac{-i}{2}\left( \glP_{\BraggOrder-2}-i\sumLoss\right)}\cos{\left(\frac{\dPhi_{\BraggOrder-2}-i\diffLoss}{2}\right)} } 
$}%checked 
&%[3,4] 
\scalebox{\MatrixScale}{$
 \begin{smallmatrix}
 \textcolor{blue}{
 \frac{1}{2} e^{\frac{-i}{2} \left(\dPhi_{\BraggOrder} +2\left(\LZphimin+\LZphiplus\right) +\glP_{\BraggOrder} + 4 \Lphase \right)} \times
 }
 \\ \textcolor{blue}{\big(e^{i \LZphimin} h_1(\sumLoss,\diffLoss) + e^{i \left(\dPhi_{\BraggOrder} +\LZphiplus\right)} h_2(\sumLoss,\diffLoss) \big)   }
 \end{smallmatrix} 
$}%checked
\Bstrut \\[5pt]\hline
  %%Row #4
%[4,1]
\Tstrut
\scalebox{\MatrixScaleSmall}{$
- i e^{i2\BraggOrder \Lphase} e^{\frac{-i}{2}\left( \glP_\BraggOrder-i\sumLoss\right)}\sin{\left(\frac{\dPhi_{\BraggOrder}-i\diffLoss}{2}\right)} $}%checked    
&%[4,2]
\scalebox{\MatrixScale}{$
\begin{smallmatrix} 
\textcolor{blue}{
\frac{-1}{2} e^{\frac{-i}{2} \left(\dPhi_{\BraggOrder-2} +\glP_{\BraggOrder-2} - 4(\BraggOrder-1) \Lphase \right)} \times
} \\
\textcolor{blue}{
\big(e^{i \LZphiplus} h_1(\sumLoss,\diffLoss)- e^{i \left(\dPhi_{\BraggOrder-2} +\LZphimin\right)} h_2(\sumLoss,\diffLoss) \big)  
}
\end{smallmatrix}
$}% checked
&%[4,3]
\scalebox{\MatrixScale}{$
\begin{smallmatrix} \textcolor{blue}{
\frac{-1}{2} e^{\frac{-i}{2} \left(\dPhi_{\BraggOrder-2} +\glP_{\BraggOrder-2} - 4 \Lphase \right)}\times}\\\textcolor{blue}{ \big( e^{i \LZphiplus}  h_1(\sumLoss,\diffLoss)+e^{i \left(\dPhi_{\BraggOrder-2} +\LZphimin\right)} h_2(\sumLoss,\diffLoss)  \big) }
\end{smallmatrix}
$}%checked 
&%[4,4]
\scalebox{\MatrixScaleSmall}{$
e^{\frac{-i}{2}\left( \glP_\BraggOrder-i\sumLoss\right)}\cos{\left(\frac{\dPhi_\BraggOrder-i\diffLoss}{2}\right)}
$}%checked   \\
        \end{array}\right)
\end{aligned}
\end{math}%  
}
\end{align}
with $h_1(\sumLoss,\diffLoss)\equiv \sqrt{1-\cosh{(\diffLoss+\sumLoss)} +  \sinh{(\diffLoss+\sumLoss)}}$ as well as $h_2(\sumLoss,\diffLoss)\equiv \sqrt{1- e^{\left(\diffLoss-\sumLoss\right)}} $, and
\begin{align} \label{SMeq:BMatrixpFirstOrd}
     \mathcal{N}(p,\Omega,\tau)=  \left(\begin{array}{@{}c|c|c|c@{}}
%%Row #1
\frac{1+ i \eta(p) \cos{\left(\dPhi_\BraggOrder / 2\right)}}{\sqrt{1+\eta^2(p)}} &\multicolumn{2}{c|}{\textcolor{blue}{\textbf{0}}} & i e^{-i2\BraggOrder \Lphase} \frac{i \eta(p) \sin{\left(\dPhi_\BraggOrder / 2\right)}}{\sqrt{1+\eta^2(p)}} \BstrutTwo\\ \hline
%%Row #2
\multirow{2}{*}{\textcolor{blue}{\textbf{0}}}&\multicolumn{2}{c|}{\multirow{2}{*}{\scalebox{1.3}
        {\textcolor{blue}{$\mathds{1}$}}}}& \multirow{2}{*}{\textcolor{blue}{\textbf{0}}}\\
%%Row #3
&\multicolumn{2}{c|}{\multirow{2}{*}{}}&\\\hline
%%Row #4
\TstrutTwo - i e^{+i2\BraggOrder \Lphase} \frac{i \eta(p) \sin{\left(\dPhi_\BraggOrder / 2\right)}}{\sqrt{1+\eta^2(p)}} &\multicolumn{2}{c|}{\textcolor{blue}{\textbf{0}}} &  \frac{1- i \eta(p) \cos{\left(\dPhi_\BraggOrder / 2\right)}}{\sqrt{1+\eta^2(p)}}  \\
        \end{array}\right). 
\end{align}
\end{subequations}
These matrices describe asymptotically the dynamics of Bragg diffraction in zeroth order of the (quasi)momentum $p$ and its first-order correction, respectively. Accordingly, the strength of the perturbation is characterized by $\eta(p) \propto p /\hbar k$ with a proportionality constant that takes into account the spectral width of the pulse and thus depends on the pulse parameters, cf. \cite{Siemss2020}. 

Except for the laser phase $\Lphase$, the choice of Bragg pulse parameters $\big\{\Omega,\tau \big\}$ in the experiment determines all quantities in Eqs.~\eqref{SMeq:BMatrixpZerothOrd} and~\eqref{SMeq:BMatrixpFirstOrd}, most of which have already been introduced in~\cite{Siemss2020}. We give a general overview of the role of each quantity and discuss how they can be calculated in the following sections. Nevertheless, we would like to emphasize that it is the analytical form of the transfer matrix~\eqref{SMeq:BMatrix}, including the coherent coupling to the dominant perturbative diffraction orders, that forms the basis of the work presented in the main article. By composing several transfer matrices, as described in Sec.~\ref{App:ScatteringFormalism}, we arrive at an analytical expression for the Bragg atom interferometer signal that gives crucial insight into its structural dependence on the individual Bragg operations. These findings motivate the formulation of the principle results in the main text in the form of Eqs.~(1) and (2), and reveal the potential to suppress unwanted interferometer paths in a straightforward manner using appropriate pulse parameters.

\subsubsection{Energetic phases:
\texorpdfstring{$\dPhi_{n}, \dPhi_{n-2},\;\glP_{n},\glP_{n-2}$}{O\_n,O\_n-2,O\_n,O\_n-2}
}
Within the adiabatic theory of Bragg diffraction, the phases $\dPhi_{n}, \dPhi_{n-2},\;\glP_{n},\glP_{n-2}$ are determined by energetic phases acquired during adiabatic dynamics. Examining Eq.~\eqref{SMeq:BMatrixpZerothOrd} and comparing it to the ideal operations in Eq.~\eqref{SMeq:TwoModeBraggMatrices} makes clear, that the phase $\dPhi_{\BraggOrder}$ determines, whether the pulse acts as a Bragg beam splitter ($\dPhi_{\mathrm{BS},\BraggOrder} = \frac{\pi}{2}$) or a Bragg mirror ($\dPhi_{\mathrm{M},\BraggOrder} = \pi$) within the subspace $\{\ket{\pm \BraggOrder\hbar k}\}$. Therefore, it is also referred to as the condition on the Bragg pulse area~\cite{Mueller2008PRA,Siemss2020}. Interestingly, various combinations of $\big\{\Omega,\tau \big\}$ satisfy this pulse area condition, which allows to trade-off the losses due to velocity-dependent Doppler effects against the strength of the unwanted diffraction orders. The corresponding parameters are described in more detail in the next two sections. 

In analogy, the phase $\dPhi_{\BraggOrder-2}$ plays a similar role within the subspace of parasitic states $\{\ket{\pm (\BraggOrder-2)\hbar k }\}$, while $\glP_{n}$ as well as $\glP_{n-2}$ are phases imprinted on the state during the Bragg pulse, cf.~\cite{Gochnauer2019,Siemss2020}. In general, all of these quantities are related to time integrals of the instantaneous eigenenergies of the lattice Hamiltonian in the sense of the adiabatic theorem, see~\cite{Siemss2020}. This allows us to find parameters highlighted in Sec.~\ref{App:Parameters}, which yield efficient Bragg mirror operations for the main modes ($\dPhi_{\mathrm{M},\BraggOrder} = \pi$) while being transparent ($\dPhi_{\BraggOrder-2}\approx0$) for states $\{\ket{\pm (\BraggOrder-2)\hbar k }\}$. This is the principle of the adapted mirror configuration discussed in the main article and illustrated in Fig.~1(B).

\subsubsection{Non-adiabatic parameters: \texorpdfstring{$\diffLoss,\sumLoss$, $\LZphimin, \LZphiplus$}{y,F,E\_-,E\_+}}\label{App:gammaIntroduction}
Parameters $\diffLoss,\sumLoss$ and $\LZphimin, \LZphiplus$ represent corrections to the adiabatic dynamic of the Bragg pulse in terms of losses from the main states $\{\ket{\pm \BraggOrder\hbar k }\}$. $\diffLoss,\sumLoss$ are related to the strength of the coupling to the spurious diffraction orders and are in fact composed of quantities that represent two distinct loss processes described by Landau-Zener physics~\cite{Siemss2020},
\begin{align}\label{SMeq:Gammacomposition}
    \sumLoss = \diffLoss_{+} + \diffLoss_{-} \qquad \text{and} \qquad \diffLoss = \diffLoss_{+} - \diffLoss_{-}.
\end{align}
The parameters $\diffLoss_\pm$ are associated with the losses within the even($+$) and odd($-$) subspaces spanned by the momentum eigenstates $\{\ket{\BraggOrder\hbar k }\pm\ket{-\BraggOrder\hbar k }\}$ and can be calculated analytically via the theory of Landau-Zener transitions for Bragg order $\BraggOrder=2$, see~\cite{Siemss2020}. As they are linked to the population of spurious diffraction orders and we are interested in high-fidelity Bragg operations with losses $<10\%$, we can assume both parameters to be small, i.e., $\diffLoss,\sumLoss \ll1$. In Sec.~\ref{App:gamma} we evaluate $\diffLoss$ for the pulse parameters considered in the main text and discuss, how $\diffLoss,\sumLoss$ can be extracted from measurements of diffracted beam splitter populations. Meanwhile, the non-adiabatic losses also imprint different transition phases, $\LZphimin$ as well as $\LZphiplus$, which are again unique to either subspace. In this work, the values for all four parameters are obtained for given pulse parameters $\big\{\Omega,\tau \big\}$ by numerically solving the Schrödinger equation as described in Sec.~\ref{App:gamma} and using the explicit form of the matrix in Eq.~\eqref{SMeq:BMatrixpZerothOrd}.  

\subsubsection{Finite velocity effects: \texorpdfstring{$\eta(p)$}{n(p)}}
The matrix $\mathcal{N}(p,\Opeak,\tau)$~\eqref{SMeq:BMatrixpFirstOrd} accounts for first-order corrections to the dynamics of the Bragg pulse due to Doppler effects most dominant for long pulse durations. The finite momentum width of the atomic ensemble, modeled here by a Gaussian distribution $g(p,\pspread) = (2\pi \pspread^2)^{-1/4}\mathrm{exp}{\left(-p^2/4\pspread^2\right)}$~\footnote{Note, that the symbol of the momentum width $\pspread$ has been adapted from $\Delta_p$ in~\cite{Siemss2020}.}, causes non-vanishing (quasi)momenta $p$ relative to the optical lattice. 
As we assume the use of ultra-cold atomic ensembles~\cite{Kovachy2015,Deppner2021} in our perturbative treatment in~\cite{Siemss2020}, we consider $p/\hbar k \ll 1$ for all momentum components. The parameter $\eta(p)$ can also be determined via instantaneous energy eigenstates of the Bragg Hamiltonian~\cite{Siemss2020}.

Figure~5 in the main text confirms good agreement between the numerical data and our analytical model with residual deviations of the expected order of magnitude. They are most evident considering the wave packet with broader momentum distribution at a longer pulse duration. Our model overestimates the strength of the velocity filtering, especially in the case of the Bragg beam splitters, cf.~\cite{Siemss2020}. This and the fact, that they are less important compared to the mirror pulse for the pulse parameters considered in this work, we neglect finite velocity effects during beam splitting in our analytical description, i.e., $N(p,\BStuple)=\mathds{1}$, in order to avoid potential artifacts. Furthermore, we neglect elements of $N(p,\Opeak,\tau)$ that would result in terms of the order $\mathcal{O}[\gamma \frac{p}{\hbar k } ]$ or higher in Eq.~\eqref{SMeq:BraggMatrix_b} , as $\gamma$ is also a small parameter. 
%%%%%%%%%%%%%% 
%%%%%% SubSection: Scattering Matrix formalism for Bragg atom interferometers
%%%%%%%%%%%%%% 
\subsection{Scattering matrix for Bragg atom interferometers}\label{App:ScatteringFormalism}
We define the scattering matrices for Bragg interferometer sequences taking into account the dominant spurious diffraction orders in this section. First, we outline the key assumptions required to extend the description of the Bragg pulse transfer matrices to the interferometric geometries. As illustrated in Fig.~1 in the main article, in Bragg interferometery a matter wave is brought into spatial superposition of at least two different momenta following separate trajectories, which can be associated with these atomic momentum modes. In the MZ geometry, e.g., the different modes are coupled by the atom-light interaction at three distinct moments in time, which we describe via Bragg transfer matrices of the form in Eqs.~\eqref{SMeq:BraggMatrix_full}. 

This picture hinges on two assumptions. First, we consider pulse durations much shorter than pulse separation time $\tau\ll T$, which is typically the case for light-pulse atom interferometers. Second, we are interested in efficient high-order Bragg pulses requiring ultra-cold atomic ensembles with momentum widths $\pspread\ll \hbar k$~\cite{Szigeti2012,Siemss2020}. The assignment of a spatial trajectory to each momentum mode $\ket*{q_j}_{j=1,2,\dots,\totalPortsGen}$  is thus unambiguous, and trajectories separated by multiple momenta $2\BraggOrder\hbar k $ can always be measured individually if we assume sufficiently time-of-flight durations. As we will discuss in the next section, the total number of modes $\totalPortsGen$ coupled by the Bragg interferometer sequence depends on both the geometry of the interferometer and the dimension of the Bragg transfer matrices in Eq.~\eqref{SMeq:BraggMatrix_full}. Under these assumptions, we represent the incoming wave packet with an average momentum $-\BraggOrder \hbar k $ in the reference frame of the optical lattice as depicted in Fig.~1 including a Gaussian momentum distribution $g(p,\pspread) $ in a momentum basis of the unique input modes $\ket*{q_j}$
\begin{align} \label{SMeq:InputState}
    \begin{split}
        \ket{\inputState(\pspread)}
        =  \int^{\hbar k/2}_{-\hbar k/2}\difp \;c_{1}(p)\; \smash{\ket*{q_{1}(p)}}_{\mathrm{in},1} = \int^{\hbar k/2}_{-\hbar k/2}\difp \;g(p,\pspread)  \ket*{\vphantom{ket{q_j}}- \BraggOrder\hbar k +p}.
    \end{split}
\end{align}
We remark that $ \ket*{\inputState(\pspread)} $ represents a single-particle state, which is sufficient at this point to describe the dynamics of Bragg interferometers in the absence of nonlinearities, e.g., due to particle-particle interactions. Later in Secs.~\ref{App:CRBounds} and \ref{App:PhaseUncertainty} we discuss the implications of the input state consisting of many uncorrelated atoms with a Poissonian distribution of total number of atoms.

The scattering matrix for arbitrary Bragg interferometer sequences, which we also expand in the basis of modes $\ket*{q_j}$,
\begin{align}\label{SMeq:ScatteringMatrixOperator}
            \mathcal{S}=\int^{\hbar k /2}_{- \hbar k /2}\difp\sum_{j,l=1}^{\totalPortsGen}\; [\mathcal{I}(p)]_{jl}\hspace{3pt}  \smash{\ket{q_j}}_{\mathrm{out},j} \hspace{1.5pt}{}_{\mathrm{in},l\!}\bra{\vphantom{ket{q_j}}q_{l}}, 
\end{align}
describes the action of the atom interferometer on the input state, 
\begin{align}\label{SMeq:OutputStateGen}
          \ket{\outputState(\phi, T,\BStuple,\Mtuple,\pspread) } = \mathcal{S}  \ket{\inputState(\pspread)}  =  \int^{\hbar k /2}_{- \hbar k /2}\difp\;g(p,\pspread)\;\sum_{j=1}^{\totalPortsGen}\; [\mathcal{I}(p)]_{j1}\hspace{3pt}  \smash{\ket{q_j}}_{\mathrm{out},j}.         
\end{align}
Here, the matrix $\mathcal{I}(p)$ is unique to a particular interferometer scheme, since it encodes the interactions of the Bragg pulses based on the individual transfer matrices in Eq.~\eqref{SMeq:BraggMatrix_b} as well as the free propagation in between.
%%%%%%%% 
%%%%%% SubSubSection: Mach-Zehnder interferometer
%%%%%%%%
\subsubsection{Application to the Mach-Zehnder interferometer}\label{App:MZScatteringMatrix}
\begin{figure}[ht]
        \includegraphics[width=0.65\textwidth]{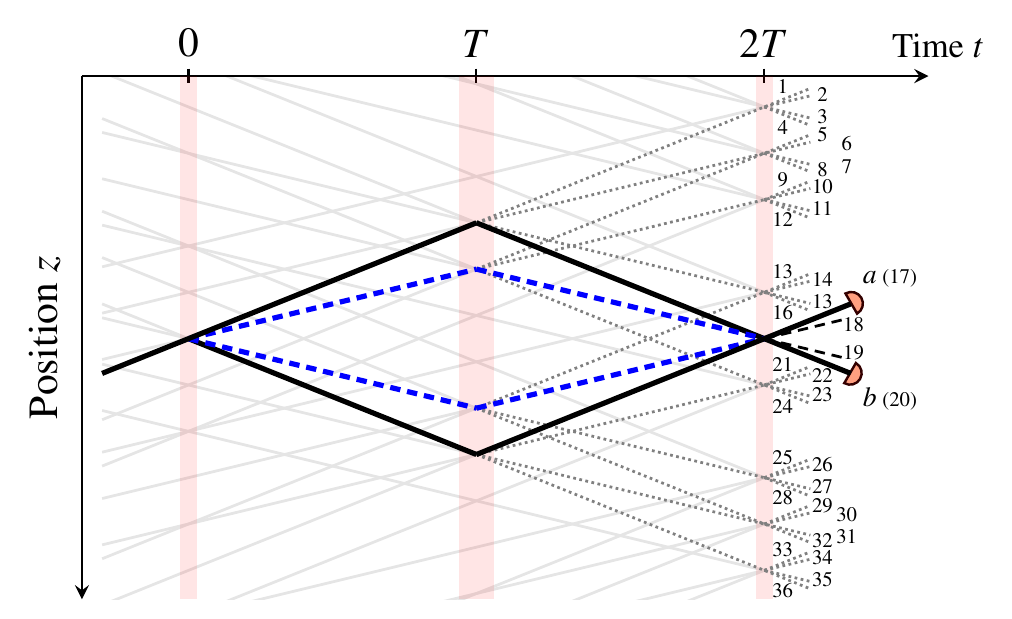}
        \caption{Space-time trajectories of the symmetric Mach-Zehnder~(MZ) interferometer considering fifth-order Bragg diffraction and including the dominant undesired diffraction order. Besides the main (solid black lines) and the parasitic (dashed blue lines) interferometry paths, cf. Fig.~1(A) in the main text, loss channels coupled by the mirror and final beam splitter interaction are displayed as well (dotted gray lines). The definition of the scattering matrix in Eq.~\eqref{SMeq:ScatteringMatrixOperator} includes the $36$ trajectories for all times $t$ even if they are unpopulated (solid light-gray lines) using the initial state $\ket*{\inputState} = c_{1}\ket*{q_{1}}_{\mathrm{in},1}$ defined in Eq.~\eqref{SMeq:InputState}. We regard coupling to output ports other than $\{a(17),18,19,b(20)\}$ as incoherent atom loss.    
}\label{SMfig:MZGeometry}
\end{figure}

In this section, we construct the scattering matrix for the MZ geometry and derive analytical expressions for the interferometer signal, assuming fifth-order Bragg diffraction pulses and considering the dominant spurious diffraction orders. These results are the basis for Eqs.~(1) and~(2) presented in the main text. Fig.~\ref{SMfig:MZGeometry} shows all $\totalPortsGen=\totalPortsMZ$ trajectories populated by the $4\times 4$-transfer matrices, based on which Eq.~\eqref{SMeq:ScatteringMatrixOperator} is defined. We note that this accounts for the dominant diffraction orders only, and that there will be further, less strongly populated parasitic interferometer paths due to diffraction beyond the state space of Eq.~\eqref{SMeq:statespace}.

%%
%%% FIG. S1 - Panel: Trajectories
%%
\FloatBarrier
 Formally, we construct the scattering matrix for the interferometer by composing $\mathcal{I}(p)$ in Eq.~\eqref{SMeq:ScatteringMatrixOperator} of individual transfer matrices representing the different elements of the sequence
\begin{align} \label{SMeq:MZMatrixMultiplication}
            \mathcal{I}_\mathrm{MZ}(p) = \mathcal{B}_{\mathrm{2.BS}}\cdot\mathcal{U}\cdot\mathcal{B}_{\mathrm{M}}\cdot\mathcal{U}\cdot\mathcal{B}_{\mathrm{1.BS}}.
\end{align}
Bragg matrices $\mathcal{B}_{\Lambda} \equiv \mathcal{B}(p,\Opeak_{\Lambda},\tspread_{\Lambda},\phi_{\mathrm{L},\Lambda})$ describe the coupling of individual trajectories due to the atom-light interaction during the two beam splitters and the mirror pulse. The subscript $\Lambda=\left\{\mathrm{1.BS},\mathrm{2.BS},\mathrm{M}\right\}$ denotes the different choices of pulse parameters $\{\Opeak,\tau,\phi_{\mathrm{L},\Lambda}\}$. Fig.~\ref{SMfig:MZGeometry} demonstrates that half of the trajectories are only populated after the final beam splitter, which is why we define these matrices as sparse,
\begin{align}\label{SMeq:MZAtomLightMatrix}
\mathcal{B}_{\Lambda}  = (\mathcal{B}_{\Lambda})_{j,l = 1,\dots,\totalPortsMZ} =
\begin{cases}
[B_{\Lambda}]_{\nu_1,\nu_2} ,& \text{if } j = (P_{\Lambda})^{s}_{\nu_1} \text{ and }  l = (P_{\Lambda})^{s}_{\nu_2} \\
1,              & \text{if } j=l \not\in P_{\Lambda} \\
0,              & \text{otherwise} . 
\end{cases}
\end{align}
$[B_{\Lambda}]_{\nu_1,\nu_2}$ denote elements of matrices in Eq.~\eqref{SMeq:BraggMatrix_b} and  $P_{\Lambda}$ is an index list of trajectories $\ket*{q_j}_j$ populated during the diffraction process. For the example shown in Fig.~\ref{SMfig:MZGeometry} and the four-dimensional state-space in Eq.~\eqref{SMeq:statespace} we find 
\begin{alignat}{4}
&P_{\mathrm{1.BS}} &&= (P_{\mathrm{1.BS}})_{\tsub{$s=1$\\$\nu=1, \dots , 4$}}&&=            \begin{cases}
9\cdot \nu ,        & \text{if } \nu>2 \\
9\cdot \nu -8,    & \text{otherwise} . 
\end{cases}&&= \{1,10,27,36\},\\%
&P_{\mathrm{M}} &&= (P_{\mathrm{M}})_{\tsub{$s=1,\dots,4$\\$\nu=1, \dots , 4$}}
&&= \nu + 4 (\nu+s-2)+            \begin{cases}
0,        & \text{if } s,\nu\leq2\\
8,        & \text{if } s,\nu>2\\
4,     & \text{otherwise} . 
\end{cases}&&= \{\{1,6,15,20\},\{5,10,19,24\},\{13,18,27,32\},\{17,22,31,36\} \}, \\
&P_{\mathrm{2.BS}} && = (P_{\mathrm{2.BS}})_{\tsub{$s=1,\dots, 9$\\$\nu=1, \cdots , 4$}}  && = \nu + 4(s-1)  &&= \{\{1,2,3,4\},\{5,6,7,8\},\dots,\{33,34,35,36\}\}.
\end{alignat} 

The matrices $\mathcal{U}$ in Eq.~\eqref{SMeq:MZMatrixMultiplication} imprint the metrological phase on the state acquired during the free propagation between pulses. For illustration, we model the phase evolution caused by a linear potential $V_g=mg\hat{z}$ acting on the atoms in between pulses and imprinting a relative phase $\BraggOrder\phi_g = 2\BraggOrder k g T^2$ between the main interferometer arms, e.g., cf.~\cite{Hogan2009}. Following \textit{Kritsotakis et al.}~\cite{Kritsotakis2018}, a state $\ket{\psi}$ evolves in this potential according to $\ket{\psi(T)} = \hat{U}(T)  \ket{\psi}$ with the unitary operator
\begin{align} \label{SMeq:FreePropagation}
    \hat{U}(T)= \mathrm{exp} \left[-\frac{iT}{\hbar}\left(\frac{\hat{\textbf{p}}^2}{2m} + mg \hat{z}\right)\right].
\end{align}
It follows from Eq.~\eqref{SMeq:FreePropagation}, that the gravitational acceleration also affects the trajectories, when the atom interferometer is operated in free fall. To avoid detrimental Doppler effects, this is typically compensated by adjusting the frequencies of the lattice light fields. In this case, experiments measure the effective acceleration $\abs{g_\mathrm{eff}}\equiv \abs{ g- a_\mathrm{Laser}} \ll \hbar k/(mT)$ using the control of $a_\mathrm{Laser}$. In this sense, we neglect the small change of the atomic trajectories caused by the operator in Eq.~\eqref{SMeq:FreePropagation} and consider only the evolution of the phase of the states $\ket*{q_j}_{j}$. With this and provided that the wave vector of the light fields is aligned parallel to the $z$ direction, we find 
\begin{align} \label{SMeq:FreepropagatedState}
    \begin{split}
    \ket{\psi(T)} = \hat{U}(T)  \sum^{\totalPortsMZ}_{j=1} c_{j}(p) \; \ket{q_{i}}_j  =  \sum^{\totalPortsMZ}_{j=1} c_{j}(p) \; e^{i\freepropPhase(g,T,q_j)}  \ket{q_{j} - mgT}_j \approx  \sum^{\totalPortsMZ}_{j=1} c_{j}(p) \; e^{i\freepropPhase(g,T,q_j)}  \ket{q_{j}}_j .
    \end{split}
\end{align}
Here, we use that $\ket*{q_j}_{j}$ are eigenstates of the momentum operator $\hat{\textbf{p}}$ and omit the suffix from $g_\mathrm{eff}$ for better readability. Consequently, we define the scattering matrix $\mathcal{U}$ in Eq.~\eqref{SMeq:MZMatrixMultiplication} describing a pure phase evolution during free propagation via 
\begin{align} \label{SMeq:FreePropScatteringMatrix}
    \begin{split}
        \mathcal{U} &= (\mathcal{U})_{j,l=1,\dots, \totalPortsMZ} =            \begin{cases}
                e^{i\freepropPhase(g,T,q_j)},        &\text{if } j =l \\
                0,     & \text{otherwise} . 
            \end{cases}\\    
    \end{split}
\end{align}
We can derive the analytical expression for the momentum- and therefore path-dependent propagation phase in  Eqs.~\eqref{SMeq:FreepropagatedState} and~\eqref{SMeq:FreePropScatteringMatrix} by following following the steps outlined in the appendix \textbf{A} of~\cite{Kritsotakis2018},
\begin{align}\label{SMeq:PropagationPhase}
     \freepropPhase(g,t,q_j) &  \equiv m \frac{g^2 t^3}{3 \hbar } - \frac{gt^2}{2\hbar } q_j - \frac{t}{2 m \hbar } (q_j - m g t)^2.
\end{align}
Finally, we are in a position to insert Eq.~\eqref{SMeq:MZMatrixMultiplication} into Eq.~\eqref{SMeq:ScatteringMatrixOperator} and can evaluate the output~\eqref{SMeq:OutputStateGen} for the symmetric MZ interferometer sequence.

%%%%%%%% 
%%%%%% SubSubSection: Signal of \BraggOrder=5 Mach-Zehnder interferometer
%%%%%%%%
\subsection{Signal of Mach-Zehnder interferometer for Bragg order  \texorpdfstring{$\BraggOrder=5$}{n=5}
}\label{App:MZSignalGen}
By means of the Bragg scattering matrix for the MZ interferometer developed in the previous section, we are able to calculate the analytical expressions for the atom numbers recorded at the individual output ports,
\begin{align}\label{SMeq:AbsoluteAtomNumbers}
    N_j(\phi) \equiv  N_j(\phi, T,\BStuple,\Mtuple,\pspread)  = \meanAtoms\int^{\hbar k/2}_{-\hbar k/2}\difp\;  \abs{\braket{\outputState}{q_j}_j}^2.
\end{align}
Since Eq.~\eqref{SMeq:OutputStateGen} describes only a single-particle output state, we have included the number of uncorrelated atoms entering the interferometer $\meanAtoms$ in the definition of $N_j(\phi)$ assuming that the time evolution of each atom is independent. A more nuanced discussion of Poissonian statistics in the initial atom number and their implications will be given further below in Sec.~\ref{App:PhaseUncertainty}.

\subsubsection{Signal in main ports 
\texorpdfstring{$\portI$}{a} and \texorpdfstring{$\portII$}{b} including parasitic interferometers}\label{App:MZSignalParasitic}
The signal in atom interferometry is recorded in terms of relative atom numbers, e.g., $\displaystyle P_{\portI(\portII)} = N_{\portI(\portII)}(\phi)/N_{\portI}(\phi) + N_{\portII}(\phi)$ in case of the main ports in Fig.~\ref{SMfig:MZGeometry}. Assuming fifth-order Bragg diffraction and including the dominant undesired diffraction order as shown in Fig.~\ref{SMfig:MZGeometry}, we obtain absolute atom numbers in the main output ports of the form
\begin{subequations} \label{SMeq:AbsoluteAtomnumbers}
\begin{align}  \label{SMeq:AbsoluteAtomnumbers_portI}
N_\portI(\phi) &= P_{0,\portI} + A_{1,\portI} \cos{\left(\phi + \varphi_{1,\portI}\right)} +  A_{2,\portI} \cos{\left(3\phi+ \varphi_{2,\portI}\right)} +  A_{3,\portI} \cos{\left(4\phi+ \varphi_{3,\portI}\right)} +  A_{4,\portI} \cos{\left(5\phi\right)} ,\\
 \label{SMeq:AbsoluteAtomnumbers_portII}
N_\portII(\phi) &= P_{0,\portII} + A_{1,\portII} \cos{\left(\phi + \varphi_{1,\portII}\right)} +  A_{2,\portII} \cos{\left(3\phi+ \varphi_{2,\portII}\right)} +  A_{3,\portII} \cos{\left(4\phi+ \varphi_{3,\portII}\right)} +  A_{4,\portII} \cos{\left(5\phi+ \varphi_{4,\portII}\right)}. 
\end{align}
\end{subequations}

Stating the explicit dependence of the amplitudes and phases in Eqs.~\eqref{SMeq:AbsoluteAtomnumbers} on the parameters in Eqs.~\eqref{SMeq:BMatrix} is not useful at this point as they are too unwieldy. Nevertheless, it is still worthwhile to discuss the phenomenology of the absolute atom numbers, which consist of several Fourier components of $\phi$. In addition to main interference term for $\BraggOrder=5$ represented by the last summands in Eqs.~\eqref{SMeq:AbsoluteAtomnumbers}, both expressions feature terms that depend on the metrological phase $\phi$ and arise from undesired interference between the main and the spurious interferometers arms during the final beam splitting pulse. 

It is also important to point out the distinct differences between both ports. In particular, the main interference term features no shift in port $\portI$ and a shift depending on the $\gamma$-parameters of the beam splitters of the form $\varphi_{4,\portII} = - \pi + 2\diffLoss + \mathcal{O}[\gamma]^3 $, in port $\portII$, cf.~\cite{Estey2015}. The occurrence of several Fourier components of $\phi$ and an asymmetric phase shifts in the symmetric MZ geometry is a consequence of the multi-port nature of the beam splitters. They are the reason, why the combined atom numbers $N_{\portI}(\phi)+N_{\portII}(\phi)$ become phase dependent, and the relative atom numbers $P_{\portI(\portII)}$ are therefore most generally described by Eq.~(1) in the main text, when including the effects of parasitic interferometer arms.

\subsubsection{Signal in main ports \texorpdfstring{$\portI$}{a} and \texorpdfstring{$\portII$}{b} without parasitic interferometers}\label{App:MZSignalnoParasitic}
In the main text, we show that using the Bragg mirror pulse to suppress the formation of the dominant parasitic interferometers can significantly simplify the signal of the MZ interferometer and discuss our parameter choice to achieve this in more detail in the next section. To assess the impact of the adapted mirror configuration on the interferometer signal in our analytical model we can simply set the corresponding matrix element in Eq.~\eqref{SMeq:BraggMatrix_b} to zero, $\abs*{B_{-(\BraggOrder-2),+(\BraggOrder-2)}}^2 = 0$. This leaves the absolute atom numbers in the main ports with only a single Fourier component stemming from interference of the main interferometer arms,
\begin{align}  \label{SMeq:AbsoluteAtomnumbersNoParasitic}
 N_\portI(\phi) =P_{0,\portI} +  A_{4,\portI} \cos{\left(5\phi\right)} \qquad \text{as well as}\qquad  N_\portII(\phi) =P_{0,\portII} +  A_{4,\portII} \cos{\left(5\phi  - \pi + 2\diffLoss  + \mathcal{O}[\diffLoss]^3 \right)} .
\end{align}
We note, that the amplitudes in both equations are generally different from the ones in Eqs.~\eqref{SMeq:AbsoluteAtomnumbers} and highlight that the asymmetric phase shift as a function of $\diffLoss$ remains as it is a result of spurious phases imprinted by the Bragg beam splitters in the symmetric MZ geometry. 

Using the expressions in Eqs.~\eqref{SMeq:AbsoluteAtomnumbersNoParasitic} to determine relative atom numbers in the main ports, we achieve the result given in Eq.~(2) of the main text. $P_{\portI(\portII)}(\phi)$ in Eq.~(2) is generally represented by infinite Fourier series, however, as we are interested in highly efficient LMT Bragg atom interferometers with small off-resonant population, we interpret $P_{\portI(\portII)}(\phi)$ as a series expansions in the open-port population $N_{\mathrm{open}}(\phi)$, such that the contributions of $A_j$ decrease with increasing index $j$. Specifically, in the limit of vanishing momentum width, $\pspread \rightarrow 0$, and given the additional assumption that $\dPhi_{\mathrm{BS},\BraggOrder}=\pi/2$ and $\dPhi_{\mathrm{M},\BraggOrder}=\pi$, we find
\begin{align}  \label{SMeq:RelativeAtomnumbersNoParasiticgamma}
\lim_{\pspread\rightarrow 0} P_{\portI(\portII)}(\phi) &= \frac{1}{2} \pm \frac{1}{8} (4-\diffLoss^2) \sin{\left( \BraggOrder\phi + \diffLoss + \frac{\pi}{2}\right)} \pm \frac{\diffLoss}{4} \sin{\left(2\left(\BraggOrder\phi + \diffLoss + \frac{\pi}{2}\right)\right)}  \pm \frac{\diffLoss^2}{8} \sin{\left(3\left((\BraggOrder\phi + \diffLoss + \frac{\pi}{2}\right)\right)} + \mathcal{O}[\diffLoss,\sumLoss]^3.
\end{align}

\section{High-fidelity pulse parameters and adapted mirror configuration}\label{App:Parameters}
\begin{figure}[ht]
        \includegraphics[width=1.0\textwidth]{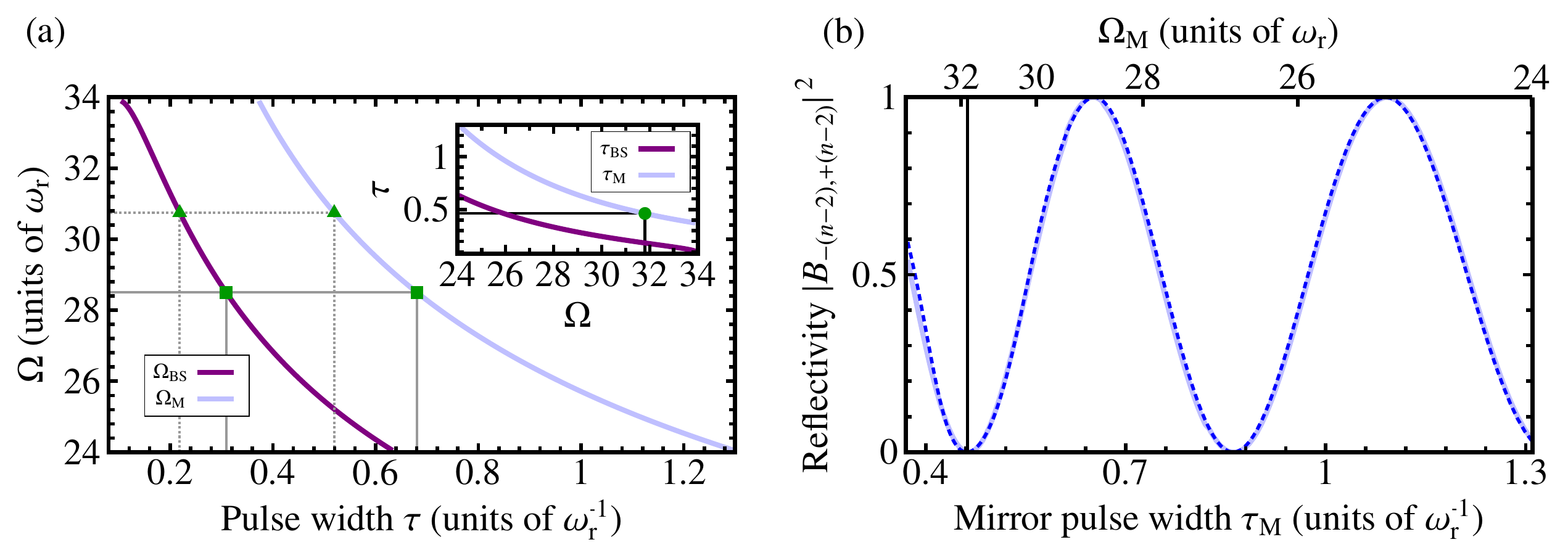}
        \caption{(a) Parameters for fifth-order Bragg beam splitter and mirror pulses used to produce the results presented in the main text. The $\BStuple$ and $\Mtuple$ tuples are determined analytically using Eq.~(51) in~\cite{Siemss2020}. The triangles (squares) as well as the dashed (solid) lines mark the parameter combinations used in Fig.~2 (Fig.~3) of the main text. We also show the mirror pulse parameters suppressing dominant parasitic interferences for $\BraggOrder=5$ in the inset (circle). (b) compares analytical (solid light blue line) and numerical (dashed blue line) solutions for the reflectivity of the Bragg mirror for the dominant parasitic order as a function of the tuples $\Mtuple$ from (a). The solid black vertical indicates the working point chosen for Fig.~4 and Fig.~5 in the main text.
}\label{SMfig:PulseParameters}
\end{figure}

In this section we give an overview of the parameters of the Gaussian pulses, $\Omega(t)=\Omega\,e^{-t^2/2\tau^2}$, used to obtain the results discussed in the main text. In general, the choice of the tuples $\BStuple$ as well as $\Mtuple$ must satisfy the aforementioned condition on the Bragg pulse area, i.e., $\dPhi_{\mathrm{BS},\BraggOrder} = \frac{\pi}{2}$ and $\dPhi_{\mathrm{M},\BraggOrder} = \pi$, respectively. They can be determined analytically by applying the adiabatic theorem~\cite{Siemss2020}. Figure~\ref{SMfig:PulseParameters}(a) showcases various combinations of pulse parameters, ranging from long to short pulse durations, that are used to obtain the results in Figs.~4 and 5 in the main article. Their range is constrained by the beam splitter diffraction losses shown in Fig.~4(c) and by the losses caused by the velocity selectivity of the mirror pulse. Moreover, the Gaussian pulse shapes are truncated in both our analytical model and the numerical simulations of the interferometer.
In the former case, we choose time intervals $t \in [-22, 22]~\omegarec^{-1}$ that reflect the asymptotic nature of our scattering theory, see~\cite{Siemss2020}. In the latter case, Gaussian pulse durations of $t\in [-10, 10]~\omegarec^{-1}$ suppress truncation effects~\cite{Mueller2008PRA}. 

Figure~\ref{SMfig:PulseParameters}(a) also indicates the parameter combinations employed in Figs.~2 and 3, and and we highlight the operating point for the Bragg mirror suppressing parasitic interferometers in the inset. As mentioned in the previous section, this operating point must satisfy the condition $\dPhi_{\mathrm{M},\BraggOrder} = \pi$ for the main diffraction order while minimize the mirror reflectivity with respect to the dominant spurious order, $\abs*{B_{-(\BraggOrder-2),+(\BraggOrder-2)}}^2$. Figure~\ref{SMfig:PulseParameters}(b) shows the analytical and numerical solutions for the reflectivity as a function of the combinations $\Mtuple$ depicted in Fig.~\ref{SMfig:PulseParameters}(a). Both solutions are in excellent agreement and allow us to identify the parameters that produce the desired suppression while being compatible with ensembles featuring finite velocity widths.

%%
%%% FIG.: Pulse parameters
%%
\FloatBarrier

%%%%%%%%%%%%%% 
%%%%%% Section: Calculation of \gamma-parameters
%%%%%%%%%%%%%% 

\section{Beam splitter population loss and calculation of \texorpdfstring{$\gamma$}{y}-parameters}
In this section we give the mathematical definition for the beam splitter diffraction losses to unwanted orders shown in Fig.~4(c) in the main text. Moreover, we outline the procedure to determine the spurious phase shift $\gamma$ included in Eqs.~(2) and~(3) in the main text. 

\subsection{Diffraction losses of fifth-order beam splitting pulse}\label{App:BSLosses}
In the idealized two-mode case, i.e., without losses as well as in the limit $\pspread \rightarrow 0$ and assuming that the parameters $\BStuple$ that satisfy the Bragg pulse area condition, the state at the output of a single Bragg beam splitting pulse,$\ket*{\outputState_{\mathrm{BS}}(\BStuple,\pspread)} = S(\BStuple)\ket*{\inputState(\pspread)}$, is in an equal superposition of modes $\ket*{\pm \BraggOrder\hbar k }$. Hence, the beam splitter loss, which we plot in Fig.~4(c) in the main text can formally be computed by evaluating 
\begin{subequations} 
\begin{align} \label{SMeq:BSLossesGen} 
\mathcal{L}_{\mathrm{BS}} &=1 -  \lim_{\pspread\rightarrow 0} \int^{\hbar k /2}_{-\hbar k/2 } \difp\;\left( \abs{\braket{-\BraggOrder\hbar k + p }{\outputState_\mathrm{BS}}}^2 + \abs{\braket{+\BraggOrder\hbar k + p }{\outputState_\mathrm{BS}}}^2 \right) \\ \label{SMeq:BSLossesAnalytical}
\overset{\dPhi_\BraggOrder = \pi/2}&{=} 1 - e^{- \sumLoss} \cosh{(\diffLoss)} , 
\end{align}
\end{subequations}
where we consider the initial state denoted in Eq.~\eqref{SMeq:InputState}. The identity in Eq.~\eqref{SMeq:BSLossesAnalytical} can be found with the help of Eq.~\eqref{SMeq:BMatrixpZerothOrd} for $\dPhi_\BraggOrder = \pi/2$ and yield a simple analytic expression for the losses. Moreover, this result makes it clear that the parameters $\sumLoss,\diffLoss$ characterize the coupling to unwanted diffraction orders in terms of Landau-Zener transitions~\cite{Siemss2020}.

Since no analytical solutions for parameters $\diffLoss,\sumLoss$ are available for the case of $\BraggOrder=5$ (in contrast to $\BraggOrder=2$, see~\cite{Siemss2020}), the losses in Fig.~4(c) are computed by inserting a numerical solution for $\ket*{\outputState_\mathrm{BS}}$ into Eq.~\eqref{SMeq:BSLossesGen}. In our case, $\ket*{\outputState_\mathrm{BS}}$ is the asymptotic solution of the time-dependent Schrödinger equation $\pdv{\ket{\psi(t)}}{t} = \hat{H} \ket{\psi(t)}$, which we solve numerically using the well-known approach of a system of ordinary differential equations in a momentum basis,e.g., see~\cite{Meystre2001,Mueller2008PRA,Szigeti2012}. We solve the dynamics in terms of the optical lattice Hamiltonian $ \hat{H}(t)= \frac{\hat{\textbf{p}}^2}{2m}+ 2\hbar \Omega(t) \cos{(k \hat{z}+\Lphase(t))}$ using Gaussian temporal pulses and parameters $\tauBS$ previously discussed in Sec.~\ref{App:Parameters}. The results of Eq.~\eqref{SMeq:BSLossesGen} are then plotted in Fig.~4(c). 

\subsection{Determination of \texorpdfstring{$\diffLoss$}{y}-parameters}\label{App:gamma}
Having introduced the $\diffLoss$-parameter in Sec.~\ref{App:gammaIntroduction}, we describe here a procedure to compute $\diffLoss$ for given beam splitter pulse parameters $\BStuple$ by numerically solving the dynamics of the atom-light interaction. Assuming calibration of the numerical simulations to experimental parameters, the following steps also allow to infer the value of $\diffLoss$ in experiments.

Equation~\eqref{SMeq:BSLossesAnalytical} implies $\diffLoss$ cannot be determined simply by measuring beam splitter populations for the input state~\eqref{SMeq:InputState}. Since $\diffLoss$ is a composite quantity describing the differential loss between the even(+) and odd(-) components of states $\ket{\pm \BraggOrder\hbar k}$ to other diffraction orders, see Sec.~\ref{App:gammaIntroduction}, one must choose different initial states 
\begin{align}\label{SMeq:gammaInputState}
    \ket{\inputState_{\pm}} \equiv  \frac{1}{\sqrt{2}} \left(\ket{-\BraggOrder\hbar k} \pm \ket{\BraggOrder\hbar k}\right),
\end{align}
again expressed in the limit $\pspread \rightarrow 0$. Accordingly, we obtain $\diffLoss$ by inserting this state into $\ket*{\outputState_{\mathrm{BS}}} = S(\BStuple)\ket*{\inputState}$ and computing the (anti)symmetric loss parameters, 
\begin{align} \label{SMeq:gammaPM}
    \diffLoss = \diffLoss_+ - \diffLoss_- \quad \text{with}\quad \diffLoss_\pm \equiv - \frac{1}{2}\; \mathrm{ln}\left(2\; \abs{\braket{-\BraggOrder\hbar k }{\inputState_{\pm}}}^2\right).
\end{align}
In this way, it is straight forward to extract $\diffLoss$ from numerical solutions of the Schrödinger equation, and in principle also from experimental measurements, provided it is possible to prepare the input states~\eqref{SMeq:gammaInputState}.

\begin{figure}[ht]
    \includegraphics[width=0.6\textwidth]{ 
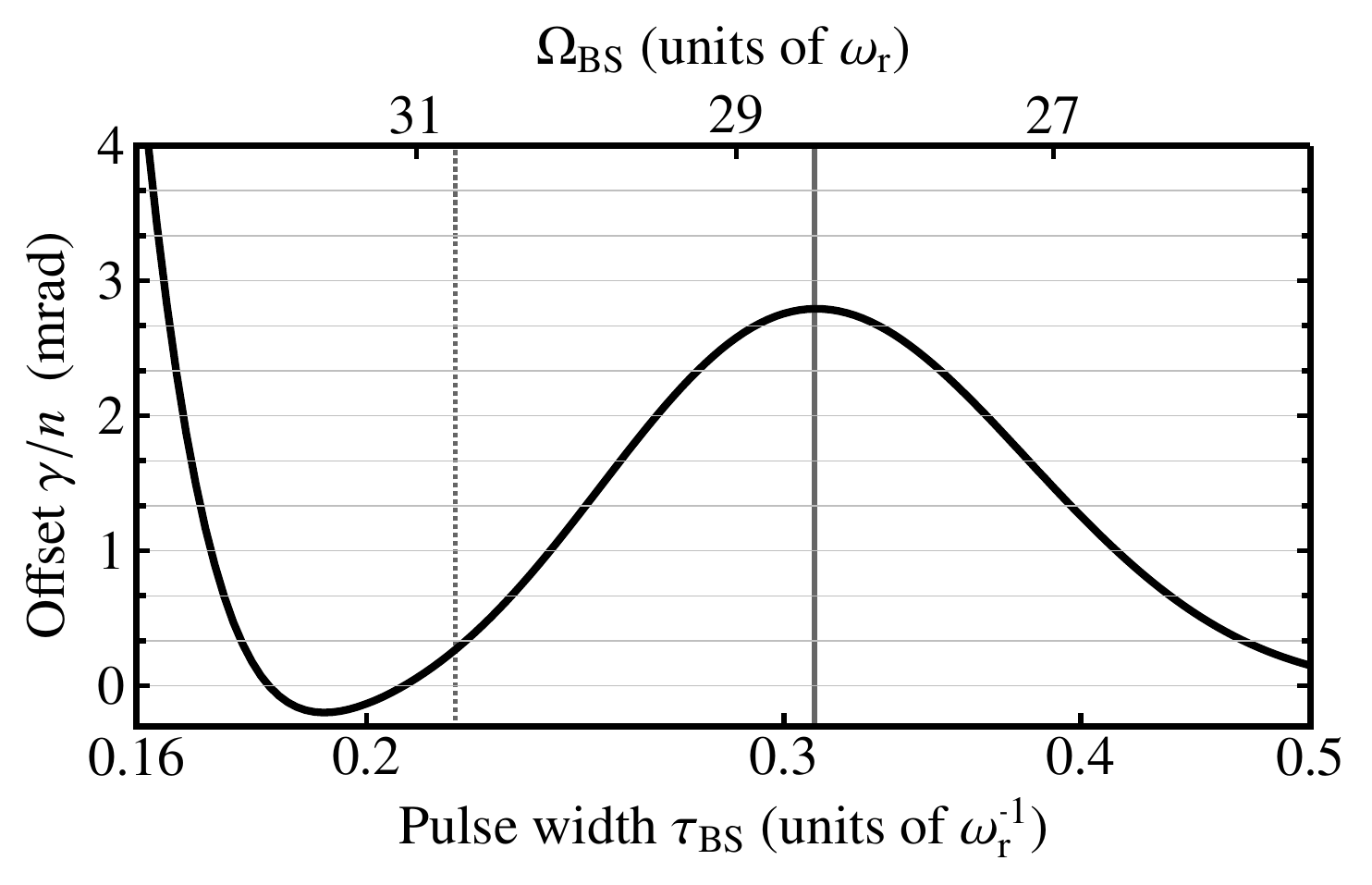}
    \caption{Evaluation of the phase offset $\diffLoss/\BraggOrder$ introduced in Eq.~(3) in the main text for a $n=5$ Bragg beam splitter. We evaluate Eq.~\eqref{SMeq:gammaPM} for parameters $\BStuple$ used in Fig.~4 in the main text. Phase shifts $\gamma/\BraggOrder$ are larger or comparable the peak-to-peak oscillations shown in Fig.~4(b) in the main text. While the maximum in $\diffLoss/\BraggOrder$ coincides with the maximum in the beam splitter losses (solid vertical line) in Fig.~4(c) in the main article, the minimum does not as it occurs for shorted pulse durations as indicated by the dotted vertical line. The fact that $\diffLoss$ can become negative is due to its composite nature explained in Sec.~\ref{App:gammaIntroduction}.
    }\label{SMfig:gamma}
\end{figure}

In Fig.~\ref{SMfig:gamma} we plot the phase offset $\diffLoss/\BraggOrder$ introduced in Eq.~(3) in the main text. Comparison with Fig.~4(c) in the main article confirms a close relationship between the $\diffLoss$-parameter and the beam splitter losses. In particular, their maxima coincide, however their minima appear for slightly different parameters. This discrepancy in combination with the fact, that $\diffLoss$ becomes negative for some parameters highlights that it is a composite quantity related to losses in two different subspaces as outlined in Sec.~\ref{App:gammaIntroduction} and indicated by Eqs.~\eqref{SMeq:gammaPM}. 

%%
%%% FIG. S2 - Panel: Statistical \gamma-values
%%
\FloatBarrier

%%%
%%%% Appendix E: Calculation of the gamma-parameters 
%%%
\section{Sensitivity bound of multi-port Bragg interferometers with uncorrelated particles}\label{App:CRBounds}
In the following, we briefly state the analytic expressions for the (Quantum) Cramér-Rao bounds~\cite{Helstrom1969} that have been evaluated and are displayed in Fig.~5 of the main text. For a detailed overview over the concepts of the quantum theory of phase estimation, we refer the reader, e.g., to the work of \textit{L. Pezzè et al.} in Ref.~\cite{Pezze2018}.    
%%%%%%%%%%%%%% 
%%%%%% SubSection: Scattering Matrix for individual Bragg pulses
%%%%%%%%%%%%%% 
\subsection{Cramér-Rao bound}\label{App:ClassicalRB}
The classical Cramér-Rao bound~(CRB) for the statistical uncertainty of estimating the based on atom number measurements discussed in main text is calculated by evaluating the Fisher information $F_\mathrm{cl}$ and leveraging its additive properties for $\meanAtoms$ uncorrelated measurement events according to
\begin{align}\label{SMeq:CRB}
        \Delta \phi_\mathrm{CRB}(\phi) &\equiv \frac{1}{\sqrt{ F_\mathrm{cl}(\phi)}}\quad \text{with}\quad  F_\mathrm{cl}(\phi)=\sum_{j=\portI,\portII}\frac{(\partial_{\phi} N_j(\phi))^2}{N_j(\phi)}\propto \meanAtoms.
\end{align}
 This expression assumes unbiased estimation and $F_\mathrm{cl}$ is calculated  for a measurement performed in the main ports $\portI,\portII$. We plot CRB in Fig.~5 in the main text using the analytical result for the absolute atom numbers in Eq.~\eqref{SMeq:AbsoluteAtomnumbers}.

\subsection{Quantum Cramér-Rao bound}\label{App:QuantumRB}
Moreover, in Fig.~5 we compare the bounds to the statistical phase estimation uncertainty using atom counting to the quantum Cramér-Rao bound~(QCRB) for arbitrary measurement operations. We restrict those measurements to be performed on the subspace of the four output ports of the final beam splitter visible in Fig.~\ref{SMfig:MZGeometry} and treat transitions outside of that as incoherent atom loss, e.g., due to velocity filtering. Formally, this can be achieved by conditioning the output state $\ket*{\outputState}$ on those trajectories, that spatially overlap with the main interferometer arms at $t=2T$. A suitable projection on the respective subspace $D \equiv \{\portI(17),18,19,\portII(20)\}$ is given by
\begin{align} \label{SMeq:OutputProjection}
        \ket{\outputStateTilde_{D}} \equiv \mathds{P}_{D}  \ket{\outputState} \quad\text{, where}\quad \mathds{P}_{D} =   \int^{\hbar k /2}_{-\hbar k/2 } \difp \hspace{3pt} \sum_{j\in D} \hspace{3pt}  \smash{\ket{q_j}}_{\!j} \hspace{1.5pt}{}_{j\!}\bra{\vphantom{ket{q_j}}q_j} .  
\end{align} %\left(\norm{\tilde{\psi}^{D}_\mathrm{out}} \right)^{-1}
After renormalizing the result to a conditional pure state,  $\ket*{\outputState_{D}} \equiv   \ket*{\outputStateTilde_{D}} / \norm{\outputStateTilde_{D}}$, we can obtain the QCRB for unbiased estimation evaluated in Fig.~5 in the main text by means of the single particle quantum Fisher information $F_\mathrm{Q}$
\begin{align}\label{SMeq:QCRB}
    \Delta \phi_\mathrm{QCRB} &\equiv  \frac{\norm{\outputStateTilde_{D}}}{\sqrt{N_\mathrm{atoms}}} \frac{1}{\sqrt{F_\mathrm{Q}}}\quad \text{with}\quad  F_\mathrm{Q}= 4\left(\braket{\outputStateDeriv_{D}}{\outputStateDeriv_{D}} - \abs{\braket{\outputStateDeriv_{D}}{\outputState_{D}}}^2\right)\;\text{, where}\; \ket{\outputStateDeriv_{D}} = \dv{\ket{\outputState_{D}}}{\phi}.
\end{align}
Note, that we have again exploited the additivity of $F_\mathrm{Q}$ for uncorrelated particles and just as one would expect, losses outside of the trajectories contained in subspace $D$ lead to an increase of statistical uncertainty due to particle loss, which is accounted for by the norm $\norm{\outputStateTilde_{D}}$ in Eq.~\eqref{SMeq:QCRB}.

%%%
%%%% Appendix C: Phase uncertainty for poissionain distributed atom number 
%%%
\section{Statistical uncertainty of phase estimate \texorpdfstring{$\statUncert$}{AO\_est}}\label{App:PhaseUncertainty}
We proceed with the derivation of the statistical uncertainty of the phase estimate $\statUncert(\phi)$ evaluated in Fig.~5 in the main text. We employ the method of moments~\cite{Pezze2018} to infer the phase $\phi$ from changes in the statistical properties of the relative atom number measurements in ports $\portI$ and $\portII$,
\begin{align}\label{SMeq:EstUncertgeneral}
   \statUncert(\phi) = \frac{\Delta \hat{P}_\portI}{\sqrt{\meanAtoms}\abs{\partial_{\phi} \expval{\hat{P}_\portI}^2 }}  = \frac{\sqrt{\expval{\hat{P}_\portI} -   \expval{\hat{P}_\portI}^2}}{\sqrt{\meanAtoms}\abs{\partial_{\phi} \expval{\hat{P}_\portI} }} \qquad\text{with}\qquad  \hat{P}_\portI = \frac{\NumOpPortI}{\NumOpPortI+\NumOpPortII} .
\end{align}
Note, that we restrict the operator $\hat{P}_\portI$ defined in Eq.~\eqref{SMeq:EstUncertgeneral} to measurements of states with non-vanishing occupation numbers in ports $\portI$ and $\portII$ respectively. 
Is becomes clear from Eq.~\eqref{SMeq:EstUncertgeneral}, that the phase uncertainty is proportional to $\meanAtoms^{-1/2}$, however, the number of uncorrelated particles entering the interferometer is fluctuating from shot to shot. When $\meanAtoms$ is statistically distributed, losses to modes $\ket*{q_j}_{\!j}$ not being detected after the final beam splitter will affect the uncertainty $\Delta \hat{P}_\portI$. We model this by assuming the initial state being prepared on the input trajectory $j=1$ as shown in Fig.~\ref{SMfig:MZGeometry} to be in a Poissonian mixture of number states $\ket{l}$,
\begin{align}\label{SMeq:StatMixInputPoisson}
\inputDensity = 
e^{- \meanAtoms} \sum^{\infty}_{l=0} \frac{(\meanAtoms)^l}{l!} \ketbra{l}{l}_{{\mathrm{in},1}}= \frac{1}{2\pi} \int^{2\pi}_{0} \diftheta\;  
\ketbra{\sqrt{\meanAtoms} e^{i \theta}}{\sqrt{\meanAtoms} e^{i \theta}}_{{\mathrm{in},1}}.
\end{align}
Here, we have suppressed all other input modes that are assumed to be vacuum modes and we have introduced the mean number of atoms $\meanAtoms$ entering the interferometer. In the second equality we have exploited the fact, that $\inputDensity$ can also be written as a phase averaged mixture of coherent states, following the work of \textit{S. Haine and M. T. Johnsson} in~\cite{Haine2009}. The states $\ket*{\sqrt{\meanAtoms} e^{i \theta}}$ in Eq.~\eqref{SMeq:StatMixInputPoisson} are Glauber coherent states~\cite{Scully1997}, $\ket{\alpha}\equiv e^{- \abs{\alpha}^2} \sum_{l} \frac{\alpha^l}{\sqrt{l!}}\ket{l}$, with amplitude $\alpha=\sqrt{\meanAtoms} e^{i \theta}$.

We point out, that when evaluating the statistics of the relative atom number measurement in Eq.~\eqref{SMeq:EstUncertgeneral} it is necessary to take into account finite velocity detuning effects and acknowledge the fact, that the experimentally determined atom numbers are averages over the finite Gaussian momentum spread $g(p,\pspread)$ of the initial state $\ket*{\psi^\mathrm{in}}$ given in Eq.~\eqref{SMeq:InputState}. We do so by expanding the annihilation and creation operators of the input modes in their respective mode functions depending on the (quasi)-momentum $p$, 
\begin{align}\label{SMeq:NumperOpBasisp}
    \hat{\portI}_\mathrm{in,1} =  \int^{\hbar k /2}_{-\hbar k/2 } \difp \hspace{3pt} g(p,\pspread)\; \hat{\Psi}_\mathrm{in,1}(p),
\end{align}
 where $\comm*{\hat{\Psi}_\mathrm{in,1}(p)}{ \hat{\Psi}^\dag_\mathrm{in,1}(\bar{p})} = \delta(p-\bar{p})$ and  $\comm{\hat{\portI}_\mathrm{in,1}}{\hat{\portI}_\mathrm{in,1}^\dagger}=1$ for the Gaussian wave packet $g(p,\pspread)$. We can describe the action of the transfer matrix in Eq.~\eqref{SMeq:MZMatrixMultiplication} on this state by first denoting the input-to-output relation of the operators in Eq.~\eqref{SMeq:NumperOpBasisp},
\begin{align} \label{SMeq:InputOutputRelationOne}
    \hat{\Psi}_{\mathrm{out},j}(p)  = \left[\mathcal{I}_\mathrm{MZ}(p)\right]_{j,1}  \hat{\Psi}_\mathrm{in,1}(p)
\end{align}
which allows us to describe the unitary evolution of the MZ interferometer in the Heisenberg picture. Note, that we have again neglected all unoccupied input modes on the right-hand sides of Eqs.~\eqref{SMeq:InputOutputRelationOne} as they will not contribute in a measurement of the occupation numbers in the output ports of the interferometer. The propagated and normalized annihilation operator for measurements in the main output port $\portI$ then reads
\begin{align} \label{SMeq:InputOutputRelationTwo}
   \hat{\portI}_{j}\equiv  \hat{\portI}_{\mathrm{out},j} = \frac{1}{w_j} \int^{\hbar k /2}_{-\hbar k/2 } \difp \hspace{3pt} g(p,\pspread)\;  \left[\mathcal{I}_\mathrm{MZ}(p)\right]_{j,1} \hat{\Psi}_\mathrm{in,1}(p),
 \end{align}  
  with a normalization factor 
  \begin{align}
   w_{j}(\phi) \equiv \left(  \int^{\hbar k /2}_{-\hbar k/2 } \difp \hspace{3pt} \abs{g(p,\pspread)}^2 \abs{\left[\mathcal{I}_\mathrm{MZ}(p)\right]_{j,1}}^2\right)^{\frac{1}{2}}.
\end{align}
In reference to Eqs.~\eqref{SMeq:EstUncertgeneral} we continue to use the shorthand $\hat{\portI}^{(\dag)}\equiv  \hat{\portI}^{(\dag)}_\mathrm{out,17} $  and $\hat{\portII}^{(\dag)}\equiv  \hat{\portI}^{(\dag)}_\mathrm{out,20} $, when referring to measurements in the main ports $\portI$ and $\portII$ as labeled in Fig.~\ref{SMfig:MZGeometry}. With the help of the input-output relation~\eqref{SMeq:InputOutputRelationTwo}, we obtain the output density operator for the input in Eqs.~\eqref{SMeq:StatMixInputPoisson},
\begin{align}\label{SMeq:StatMixOutputCoherent}
\outputDensity &= \mathcal{S}_\mathrm{MZ}\; \inputDensity\; \mathcal{S}^\dag_\mathrm{MZ}=\frac{1}{2\pi} \int^{2\pi}_{0} \diftheta\;  \bigotimes^{\totalPortsMZ}_{j=1}
\ketbra{ w_{j}(\phi) \sqrt{\meanAtoms} e^{i \theta}}{w_{j}(\phi) \sqrt{\meanAtoms} e^{i \theta}}_{{\mathrm{out},j}},
\end{align}
where the normalizations $w_{j}(\phi)\equiv w_{j}(\phi, T,\BStuple,\Mtuple,\pspread)$ crucially depend on the matrix representation of the MZ interferometer in Eq.~\eqref{SMeq:MZMatrixMultiplication} as they contain dependencies of the output on the experimental parameters and in particular the interferometric phase $\phi$. We calculate the statistics of the relative atom number measurements in the main ports~\eqref{SMeq:InputOutputRelationTwo} with respect to this output state by reformulating $\hat{P}_\portI$ in Eq.~\eqref{SMeq:EstUncertgeneral} in the limit of non-vanishing occupation numbers
\begin{align}\label{SMeq:RelativeNumberOperator}
 \hat{P}_\portI = \frac{\NumOpPortI}{\NumOpPortI + \NumOpPortII}  = \int^{1}_{0} \difx \; \big[\frac{\partial}{\partial x} x^{\NumOpPortI}\big] \otimes \big[x^{\NumOpPortII}\big], 
\end{align}
and similarly for port $\portII$. With this and Eq.~\eqref{SMeq:StatMixOutputCoherent} we find
\begin{align}\label{SMeq:MeanRealAtomNumbersDeriv}
\begin{split}
\expval{\hat{P}_\portI} & =\mathrm{tr}(\outputDensity \hat{P}_\portI)= \int^{1}_{0} \difx \; \left( \frac{\partial}{\partial x} \expval{x^{\NumOpPortI}}{ w_{\portI}(\phi) \sqrt{\meanAtoms} } \right)  \expval{x^{\NumOpPortII}}{ w_{\portII}(\phi) \sqrt{\meanAtoms} } \\
& = \meanAtoms \abs*{w_{\portI}(\phi)}^2  \;e^{-\meanAtoms\;(\abs*{w_{\portI}(\phi)}^2 +\abs*{w_{\portII}(\phi)}^2)}  \;\int^{1}_{0} \difx  \left( \;e^{x\, \meanAtoms \;(\abs*{w_{\portI}(\phi)}^2 + \abs*{w_{\portII}(\phi)}^2) }  \right),
\end{split}
\end{align}
where we have made use of the relations $\abs{\braket{\alpha}{0}}^2 = e^{-\abs{\alpha}^2}$ as well as $\expval{x^{\numOPportI}}{\alpha}  = e^{-\abs{\alpha}^2} e^{x\abs{\alpha}^2}$ in the last line. Identifying the absolute atom numbers as defined in Eq.~\eqref{SMeq:AbsoluteAtomNumbers},
\begin{align}
 N_j(\phi) = \meanAtoms  \abs*{w_{j}(\phi)}^2 = \meanAtoms  \int^{\hbar k /2}_{-\hbar k/2 } \difp  \hspace{3pt} \abs{g(p,\pspread)}^2 \; \abs{[\mathcal{I}^{*}_\mathrm{MZ}(p)]_{j,1}}^2\;, 
\end{align}
and taking the limits $\meanAtoms \abs*{w_{\portI}(\phi)}^2,\meanAtoms\abs*{w_{\portII}(\phi)}^2 \gg 1$, the statistical quantities with respect to $\hat{P}_\portI$ in Eq.~\eqref{SMeq:NumperOpBasisp} simplify to 
\begin{align}\label{SMeq:StatisticsRealAtomNumbers}
    \expval{\hat{P}_\portI} \approx \frac{ N_\portI(\phi) }{N_\portI(\phi)+N_\portII(\phi)} = P_\portI(\phi)\qquad \text{and} \qquad  \expval{\hat{P}^2_\portI} & \approx P^2_\portI(\phi) + \frac{P_\portI(\phi) \cdot  P_\portII(\phi)}{N_{\portI}(\phi)+N_{\portII}(\phi)}.
\end{align}
By inserting both results into Eq.~\eqref{SMeq:EstUncertgeneral} we arrive at an expression for the statistical uncertainty of phase estimates performed in port $\portI$, 
\begin{align}\label{SMeq:EstUncertEvaluated}
    \statUncert(\phi) =\frac{\sqrt{\expval{\hat{P}^2_\portI} -   \expval{\hat{P}_\portI}^2}}{\abs{\partial_{\phi} \expval{\hat{P}_\portI} }}  = \sqrt{\frac{P_\portI(\phi) \cdot  P_\portII(\phi)}{N_{\portI}(\phi)+N_{\portII}(\phi)}}\; \frac{1}{\abs{\partial_{\phi} P_\portI(\phi) }} ,
\end{align}
and in port $\portII$ respectively. The results $\statUncert(\phi)$  shown in Fig.~5 are obtained using Eq.~\eqref{SMeq:EstUncertEvaluated} based on numerical simulations of the MZ interferometer as described in the main text. To that end, the model $P_\portI(\phi)$ introduced in Eq.~(2) is fitted to numerical MZ signals and also atom numbers $N_{\portI,\portII}(\phi)$ are determined in the simulations. 

It is important to point out, that both the CRB in Eq.~\eqref{SMeq:CRB} as well as the statistical uncertainty of the phase estimate in Eq.~\eqref{SMeq:EstUncertEvaluated} are local quantities that depend on the value of $\phi$. In Fig.~5 in the main text, we evaluate both quantities at the mid fringe position $\localPhase = \frac{3\pi}{2} \frac{1}{5}$ for optimal performance. The presence of several Fourier components of and their harmonics in the signal of the MZ interferometer causes projection noise to be worse at neighbouring mid fringe positions, e.g., $\localPhase = \frac{1\pi}{2} \frac{1}{5}$. Here, mainly a reduction in the slope $\abs{\partial_{\phi} P_\portI(\phi)}$ is responsible for a several percent higher projection noise and projection noise limit for pulse parameters with non-negligeable losses to spurious diffraction orders.

%%%%%%%%%%%%%%%%%%%%%%%%%%%%%%%%%%%%%%%%%%%%%%%%%%%%%%%%%%%%%%%%
%%%     Bibliography
%%%%%%%%%%%%%%%%%%%%%%%%%%%%%%%%%%%%%%%%%%%%%%%%%%%%%%%%%%%%%%%% 
\bibliographystyle{apsrev4-1_custom}
\bibliography{bibliography}